\documentclass[preprintnumbers,aps,twocolumn,prd]{revtex4}
\usepackage{graphicx}
\usepackage{color}
\usepackage{latexsym}

\newcommand{\be}{\begin{equation}}
\newcommand{\ba}{\begin{eqnarray}}
\newcommand{\ee}{\end{equation}}
\newcommand{\ea}{\end{eqnarray}}

\newcommand{\barr}{\begin{array}}
\newcommand{\ear}{\end{array}}

\begin{document}

\preprint{CERN-PH-TH/2005-096}

\preprint{ FTPI-MINN-05/16}

\preprint{ UMN-TH-2403/05}  

\vspace{2cm}

\title{Moduli evolution in the presence of flux compactifications}

\author{Tiago Barreiro}
\email{tiagobarreiro@fisica.ist.utl.pt}
\affiliation{Centro Multidisciplinar de Astrof\'{\i}sica (CENTRA), Instituto Superior T\'ecnico,
Universidade T\'ecnica de Lisboa, Av. Rovisco Pais 1, 1049-001 Lisboa
PORTUGAL }
\author{Beatriz de Carlos}
\email{B.de-Carlos@sussex.ac.uk}
\altaffiliation[also at ]{Department of Physics CERN, Theory Division, 1211 Geneva 23, Switzerland}
\affiliation{Department of Physics and Astronomy, University of Sussex, Brighton BN1 9QJ, UK}
\author{Ed Copeland}
\email{ed.copeland@nottingham.ac.uk}
\affiliation{The School of Physics and Astronomy, University of Nottingham, University Park, Nottingham NG7 2RD, UK}
\author{Nelson J. Nunes}
\email{nunes@physics.umn.edu}
\affiliation{School of Physics and Astronomy, University of Minnesota, 116 Church Street S.E., Minneapolis, Minnesota 55455, USA}
\date{\today}
\begin{abstract}
We study the cosmological evolution of the volume moduli in a class of recently proposed Inflationary Universe models arising out of Type IIB string theory~\cite{Kachru:2003sx}, where a number of the moduli fields have been stabilised through flux compactifications \cite{Kachru:2003aw}. Developing an approach initially introduced in~\cite{Barreiro:1998aj} we show, in agreement with \cite{Brustein:2004jp}, how the presence of extra sources of matter act so as to provide additional friction, slowing the modulus field as it evolves down its potential, thereby vastly increasing the region of parameter space which leads to the eventual stabilisation of these fields. Extending the case to include both the real and imaginary parts of the volume modulus, we show how the parameter space of inital conditions is modified and comment on the impact for these inflationary models arising out of flux type compactifications.
\end{abstract}

\maketitle

\section{INTRODUCTION}

String theory has associated with it many moduli fields, scalar fields whose presence are important in determining quantities such as the size of the internal dimensions, the gauge coupling constants, even the strength of the gravitational interactions. One of the problems that has faced physicists attempting to develop cosmology within string theory has been how to deal with these fields -- basically they never want to stop evolving. In most models that have been developed to date, the moduli initially have no associated potential with which to trap them, the flat directions mean that there is nothing to prevent them from evolving for ever, leading to time varying coupling constants, decompactification of the internal dimensions and other nightmares. Non trivial potentials have been shown to emerge when non-perturbative features are included, such as gaugino condensates~\cite{gaugino}, but even then the local minima associated with these models are generally unstable in that the barrier height protecting it from the runaway solution is small compared to any other natural scale in the problem. Moreover, the value of the potential at its minima is negative, implying an Anti de Sitter Space solution~\cite{racetrack}. This issue has plagued string models for many years. Initially, in the context of weakly coupled heterotic string theory, this was pointed out by Dine and Seiberg~\cite{dine-seiberg}. In the context of cosmology Brustein and Steinhardt readdressed the issue, pointing out how difficult it would be to stabilise typical moduli fields emerging in these models because of the combination of the steepness of the potentials associated with the moduli, and the small size of the potential barrier separating the fields from the runaway regime~\cite{brustein-steinhardt}. Earlier, in ref.~\cite{Kaloper:1991mq}, Kaloper and Olive had
studied the effect including a background of radiation in the equations that govern the dilaton evolution.
In particular, they found that the dilaton can stop rolling if either its potential has a SUSY breaking minimum or radiation is dominating the energy density of the Universe.  In~\cite{Barreiro:1998aj} this
idea was further developed, in particular it was pointed out that, in the presence of an extra fluid component dominating the energy density, for example radiation, the effect of the fluid was to increase the effective friction experienced by the moduli field as it evolved down its steep potential. For a wide range of initial conditions of the field it was demonstrated that it would be slowed down, and its energy density would tend towards a scaling solution where it followed the evolution of the background radiation energy density. Moreover, it would then simply fall into the minimum of the potential and be stabilised there, even though the height of the potential barrier was so low. This analysis, which was initially used to demonstrate how the dilaton could be stabilised in weakly coupled heterotic string theory, was later extended to the case of heterotic M-theory, where  including the evolution of more than one modulus field, it became evident how important the coupling between the various moduli fields could be in determining the basin of attraction into the true minimum of the potential~\cite{heterotic-m-theory}. Other attempts at stabilising the moduli fields in these models included adding specific temperature corrections to the moduli fields, in order to extend the range of initial values of the fields which would lead to stabilisation~\cite{huey}, although this met with limited success~\cite{heterotic-m-theory}.

Over the past few years  there has been a renewed interest in the issue of moduli evolution.  Giddings et al demonstrated, in the context of Type IIB string theory compactified on a Calabi-Yau manifold, how it is  possible to use fluxes to stabilise all but the K\"ahler moduli, in particular the overall volume modulus~\cite{GKP}. This result was used to develop inflation models in string theory where brane inflation was embedded into one of these stable compactified models~\cite{Kachru:2003aw,Kachru:2003sx}. However in~\cite{Kachru:2003sx} it was realised that there was a price to pay, and a degree of fine tuning was required in these models in order to obtain sufficient inflation. The particular problem they encountered was that, as the compactification volume modulus was being stabilised, the effect was to modify the inflaton potential, rendering it too steep for inflation to last for a sufficient length of time. The issue of stabilisation had raised its ugly head again.  Recently Brustein et al~\cite{Brustein:2004jp} returned to the idea of using sources as a way of increasing the friction and thereby avoiding the runaway problem alluded to earlier. They did it in the context of a toy model with a single scalar field (the volume modulus) and scalar potential of the form suggested in~\cite{Kachru:2003sx}. They argued that the key reason they could stabilise the modulus was because of the way this field quickly entered a regime where its energy density was dominated by its kinetic energy. It therefore lost its energy far quicker than any other source, allowing the other sources (such as radiation) to eventually catch up and take over, slowing the field down even more and allowing it to come to rest at or near the local minimum of its potential. In this paper we extend the investigation into these models by allowing for the fact that the volume modulus is a complex field with a real and imaginary component. We therefore look at the system where more than one moduli field has to be dynamically stabilised in the presence of extra sources. We reproduce the results of~\cite{Brustein:2004jp} in the appropriate limits, but show that there are wider ranges of initial conditions which can lead to stability if we include the possibility of the moduli fields entering a period of scaling, where their energy density mimics that of the background energy density, more in the spirit of ~\cite{Barreiro:1998aj,heterotic-m-theory}. A related approach to the stabilisation issue has recently been investigated in refs.~\cite{Dimopoulos:2002hm,Kaloper:2004yj,  Easson:2005ug,Berndsen:2005qq}, where the authors have considered the presence of a gas of wrapped branes on the dynamics of some of the moduli fields. 

The layout of this paper is as follows. In section~\ref{sec:setup} we introduce the system of equations we will be investigating in terms of a general moduli potential, and briefly describe the nature of the attractor solutions when exponential terms are present in the scalar potential.  The particular example due to Kachru et al \cite{Kachru:2003sx} is analysed in section~\ref{sec:KKLT model} and a related but different example due to Kallosh and Linde \cite{Kallosh:2004yh} is examined in section~\ref{sec:KLmodel}. Finally we conclude in section~\ref{sec:conclusions}.


\section{BACKGROUND EQUATIONS OF MOTION} 
\label{sec:setup}

The class of models that we will be investigating can in general be  described by an N=1, d=4  effective Supergravity (SUGRA) theory, where d is the number of non-compact spacetime dimensions. 
In this case the four-dimensional N=1 SUGRA action is of the form
\begin{equation}
S=-\int \sqrt{-g} \left(\frac{1}{2\kappa_P^2}R+K_{i\bar j}\partial_{\mu}\Phi ^{i}\partial^{\mu}{\bar{\Phi}^{\bar{j}}}+V\right)d^4x \;\;,
\label{eq:action}
\end{equation}
where $ K_{i\bar j}=\frac{\partial^{2}K}{\partial\Phi^{i}\partial\bar\Phi^{\bar j}}$
is the K\"ahler metric; $\Phi^{i}$ are complex chiral superfields; V($\Phi$) is the scalar
potential and $\kappa_P$ is the 4-dimensional Newton constant which, in
terms of higher dimensional quantities, can be expressed as
\begin{equation}
\kappa_{P}^2=\frac{\kappa^2}{2\pi\rho v}=8\pi G_N \;\;,
\label{eq:kappa}
\end{equation}
where $\kappa$ is the 11-dimensional Newton constant, $v$ is the volume of the six dimensional compact manifold, and $\rho$ is the radius of the eleventh spatial dimension.

The impact of the non-perturbative and flux effects considered in~\cite{Kachru:2003aw,Kachru:2003sx}  is to  induce an effective scalar potential for the volume moduli whose most general form for 4-dimensional $N=1$ SUGRA is 
\begin{equation}
V =
e^{K}(K^{i\overline{j}}D_iW\overline{D_{j}W}-3W\overline{W}) \;\;.
\label{eq:potential}
\end{equation}
where $K^{i\overline{j}}$ is the inverse K\"ahler metric and
  $D_iW=\partial_{i}W+\frac{\partial K}{\partial \Phi^{i}}W$ is the
  K\"ahler covariant derivative acting on the superpotential.

The equations of motion follow from the variation of the action (\ref{eq:action}). For simplicity we  consider the case of homogeneous, time-dependent, fields in a spatially flat
 Friedmann-Robertson-Walker space time background. Given this ansatz, we find the following
equations of motion for the complex superfields
\begin{equation}
\ddot{\Phi}^{i}+3H\dot{\Phi}^i+\Gamma^{i}_{jk}\dot{\Phi}^{j}\\
\dot{\Phi}^{k}+K^{i\bar j}\partial_{\overline{j}}{V}=0 \;\;,
\label{eq:fulleom}
\end{equation}
where $\Phi^i$ are the relevant moduli, $\dot{\Phi}^i=\partial{\Phi}^i/\partial{t}$, $\partial_{\overline{j}}V=\partial{V}/\partial{\overline{\Phi}^{\overline{j}}}$,
and the connection on the K\"ahler manifold has the form
\begin{equation}
\Gamma^{n}_{ij}=K^{n\bar l}\frac{\partial K_{j\bar l}}{\partial \Phi^{i}} \; .
\label{eq:gamma}
\end{equation}

\noindent In addition, we obtain the Friedman
equation for the Hubble factor 
$H=\dot{a}/a$, where
$a(t)$ is the scale factor of the Universe,
\begin{equation}
3H^2=\kappa_P^2 (\rho_{\Phi}+\rho_b) = \kappa_P^2(
K_{i\bar j}\dot{\Phi}^{i}\dot{\Phi}^{\bar j}+V+\rho_b) \;\;,
\label{eq:fullfriedman}
\end{equation}
with $\rho_\Phi$ and $\rho_b$ are the energy density of the evolving moduli fields and background fluid respectively. The dynamics  of the latter is given in terms of the scale factor and its background equation of state, $\gamma-1 \equiv  p_b/\rho_b$, where $p_b$ is the pressure of the fluid,
\begin{equation}
\rho_b=\rho_{b0}/a^{3\gamma} \;.
\label{eq:rho}
\end{equation}
In what follows we set $\kappa_P^2 = 1$.

It is worth splitting the equations
of motion for the complex chiral superfields into those for their real and imaginary parts
\begin{equation}
\ddot{\phi}^{i}_{R}+3H\dot{\phi}^{i}_{R}+\Gamma^{i}_{jk}(\dot{\phi}^{j}_R
\dot{\phi}^{k}_R-\dot{\phi}^{j}_I\dot{\phi}^{k}_I)+\frac{1}{2}K^{i\bar j}\partial_{j_R}V=0 \;\;,
\label{eq:reom}
\end{equation}
\begin{equation}
\ddot{\phi}^{i}_I+3H\dot{\phi}^i_I+\Gamma^{i}_{jk}(\dot{\phi}^{j}_I\\
\dot{\phi}^{k}_R+\dot{\phi}^{j}_R\dot{\phi}^{k}_I)+\frac{1}{2}K^{i\bar j}\partial_{j_I}V=0 \;\;,
\label{eq:ieom}
\end{equation}
 where now $\phi^{i}_{R}$ ($\phi^i_I$) refers to the real 
 (imaginary) part of the scalar fields and  $\partial_{j_R}$
($\partial_{j_I}$) are
used to denote the derivative of the potential with respect
to the real (imaginary) parts of the fields respectively.


\section{KKLT model}
\label{sec:KKLT model}
The possibility of finding  de Sitter vacua in string theory  with a stabilized volume modulus, $\sigma$,  was put forward in ref.~\cite{Kachru:2003aw}, and has been widely adopted in subsequent work. The key ingredient was to consider the combination of non perturbative effects and an additional flux term in the superpotential
\be
W = W_0 + A {\rm e}^{-\alpha \sigma} \:\:,
\label{superpot}
\ee
which, together with the usual K\"ahler potential
\be 
K = -3 {\rm ln} (\sigma+\bar{\sigma}) \;\;,
\ee
defines the F-part of the SUGRA potential, see eq.~(\ref{eq:potential}). It has been known for many
years now that, in this context, it is possible to stabilize $\sigma$, although giving rise to an Anti de Sitter (AdS)  vacuum. As pointed out in ref.~\cite{Kachru:2003aw}, if we include contributions from either anti-D3 or D7 branes, an additional D-type term is generated, of the form
\be
V_D = \frac{C}{\sigma_r^3} \;\;,
\label{Dpot}
\ee
where we write $\sigma=\sigma_r+i \sigma_i$.
By suitably tuning the value of $C$ one can move to a dS -or even Minkowski-  vacuum.

In this section we are interested in studying the cosmological evolution of
the field $\sigma$ as it rolls towards its minimum. Previous results addressing the same 
issue were published in ref.~\cite{Brustein:2004jp}, where only the evolution of the real part 
of $\sigma$, $\sigma_r$, was considered. Here we would like to extend this to study the behaviour of both $\sigma_r$ and $\sigma_i$.

\begin{figure}[!htb]
\includegraphics[width=8cm]{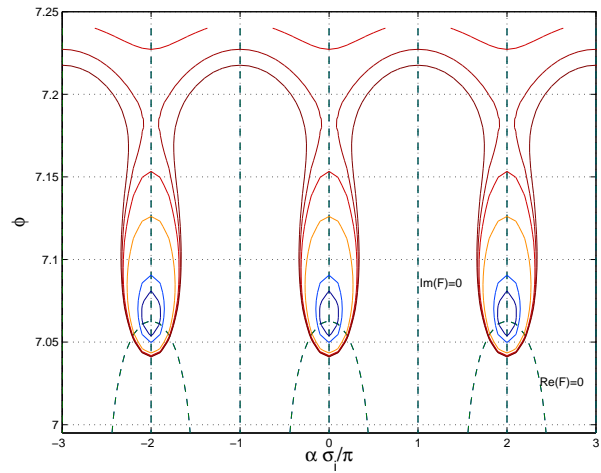}
\caption[contB]{\label{contB} Contour plot of the scalar potential, given by eq.~(\ref{potB}), in the
($\phi$, $\alpha \sigma_i/\pi$) plane (solid lines). The values of the parameters are taken from ref.~\cite{Brustein:2004jp}, see text, and the dashed lines correspond to the supersymmetry-preserving conditions, $F_{\sigma}=0$, where $F_{\sigma} \equiv K_{\sigma} W+W_{\sigma}$.}
\end{figure}

The system of differential equations that one has to solve comes from 
eqs.~(\ref{eq:reom},\ref{eq:ieom}) being applied to the present model, 
along with the evolution equation for the
background fluid. Altogether, we have
\begin{eqnarray}
\ddot{\sigma_r} + 3H\dot{\sigma_r} - \frac{1}{\sigma_r} (\dot{\sigma_r}^2-\dot{\sigma_i}^2) + 
\frac{2\sigma_r^2}{3} \partial_{\sigma_r} V & = & 0 \nonumber \;\;,\\
\ddot{\sigma_i} + 3H\dot{\sigma_i} - \frac{2}{\sigma_r} \dot{\sigma_r}\dot{\sigma_i} +
\frac{2\sigma_r^2}{3} \partial_{\sigma_i} V & = & 0 \label{evKKLT} \;\;,\\
\dot{\rho_{b}} + 3H\gamma\rho_{b} &=& 0\nonumber \;\;,
\end{eqnarray}
subject to the Friedman constraint, see eq.~(\ref{eq:fullfriedman}),
\be
3 H^2 = \frac{3}{4\sigma_r^2} (\dot{\sigma_r}^2+\dot{\sigma_i}^2) + V + \rho_{b} \;\;.
\ee
The scalar potential acquires a relatively simple form when written in terms of both real and imaginary parts of $\sigma$,
\begin{eqnarray}
V & = & \frac{\alpha A {\rm e}^{-\alpha \sigma_r}}{2\sigma_r^2} \left[ A  \left( 1+\frac{\alpha \sigma_r}{3} \right) {\rm e}^{-\alpha \sigma_r} + W_0 {\rm cos}(\alpha \sigma_i) \right] \nonumber \\
& + & \frac{C}{\sigma_r^3} \;\;. \label{potB}
\end{eqnarray}
Given the above expression it is easy to see that the potential has an extremum in $\sigma_i$ for $\alpha \sigma_i= n \pi$, with $n$ an integer. Depending on the sign of $W_0 {\rm cos}(\alpha\sigma_i)$ this can be either a maximum or a minimum. For example, in figure~\ref{contB} we show
a contour plot of this scalar potential, $V$,  in the ($\phi$,$\sigma_i$) plane, for the same values used
in ref.~\cite{Brustein:2004jp}, namely $A=1.0$, $\alpha=0.1$, $C=3 \times 10^{-26}$, and $W_0$ negative (with ${\rm cos}(\alpha \sigma_i)=1$) and such that the minimum at $\sigma_i=0$ is  supersymmetric (see below). Following that same reference we work with the canonically normalized field $\phi = \sqrt{3/2} \ln \sigma_r$, instead of $\sigma_r$ itself.

It is also worth mentioning the supersymmetric character of the minima shown in this plot. We have
included the two supersymmetry-preserving conditions, ${\rm Re} (F_{\sigma})={\rm Im} (F_{\sigma})=0$,
and it can be clearly seen how both meet at the minima.  Those correspond to $\phi=7.06$ and even
values of $\alpha\sigma_i/\pi$.

\subsection{One field evolution}
\label{sec:1field}
In order to discuss the results of ref.~\cite{Brustein:2004jp},
we will first set the potential at a minimum in $\sigma_i$ and solve only the first and third equations
of the system (\ref{evKKLT}). This would correspond to taking the slice $\alpha \sigma_i=0$ in figure~1, and considering  only the evolution along $\phi$. As for the initial conditions, we set $\dot{\sigma_r}_0=0$, and we choose values for $\sigma_{r0}$ and the fractional energy density in the background fluid, $\Omega_{b} \equiv \kappa_P^2 {\rho_{b}}_0/3H_0^2$. We can then calculate the value of $\rho_{b0}$, using  the relation
\be
{\rho_{b}}_0 = V(\sigma_{r0}) \frac{\Omega_{b}}{1-\Omega_{b}} \;\;.
\ee
In this section, the results are shown in terms of the initial abundance $\Omega_b$ which will allow us to compare them with those in ref.~\cite{Brustein:2004jp}, where $\Omega_b = 0.5$.

\begin{figure}[!hb]
\includegraphics[width=8cm]{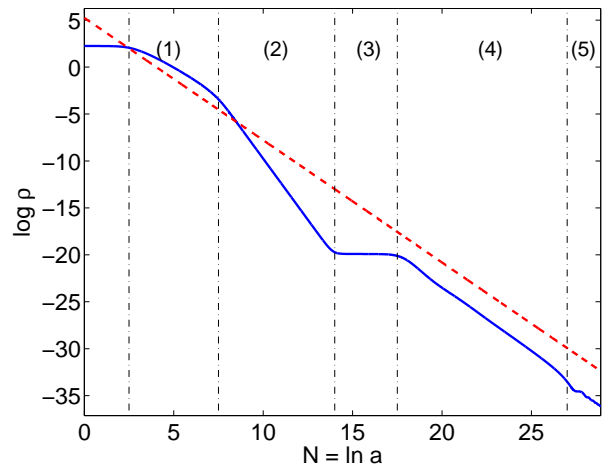}
\caption[fig20]{\label{fig20} Evolution of the energy densities of the
field $\phi$ (solid line)
and the background (dashed line) with the logarithm of the scale factor. The various epochs of the evolution, numbered from 1 to 5, are described in the text. We have taken
$\phi_0 = -5$, $\gamma = 1$ and  $\Omega_m = 0.99$. The parameters of the scalar potential are the same as used in ref.~\cite{Brustein:2004jp},
namely $A=1.0$, $\alpha=0.1$, $C=3 \times 10^{-26}$. }
\end{figure}

In general one can identify up to five regions in the evolution of a scalar field with these types of potentials. In figure~\ref{fig20} we show a typical evolution going through the five regions although, of course, not all initial conditions will give rise to an evolution that will go through all of them.

First, if the energy density of the background dominates, the field is effectively frozen in its evolution. This can be seen in the plot before region 1, with an evolution similar to region 3. When the energy density in the background becomes comparable
(region 1), the field starts rolling down its potential and eventually dominates the dynamics of the Universe. This happens when the potential is very shallow and is called the scalar field dominated solution. Eventually the potential becomes very steep, leading the evolution to a kinetically dominated solution (region 2), which ends when the frictional term in the equation of motion becomes dominant. The length of this period of "kination" is related to the value of the ratio $\rho_b/\rho_{\phi}$ at the end of region 1.
In region 3 the field is effectively frozen in the potential. The field restarts rolling down the potential once the background energy density has decayed to a value such that the frictional and potential terms in the equation of motion balance each other. At this stage the field evolves with a nearly constant ratio between kinetic and potential energy. This is called the scaling (or tracker) solution (region 4). Finally, if this ratio is sufficiently small the field does not possess enough kinetic energy to roll over the potential barrier and gets traped at the minimum (region 5).
Details of all these phases of the evolution are explained in Appendix~A.

In figure~\ref{BAM} we present the initial values of the variables ($\phi$,$\Omega_r$) for which there
is stabilization of the field at the minimum of its potential in the presence of radiation\footnote{We will use the subscript $\,_r$ when referring to a radiation background ($\gamma=4/3$), and the subscript $\,_m$ when referring to a matter background ($\gamma=0$).}, i.e. $\gamma=4/3$.
As expected, the smaller the initial background fraction (parametrized by $\Omega_{r}$) is, the
smaller the region of allowed initial values of $\phi$ which lead to a late time stabilisation of the field at the minimum. The solid line along $\Omega_r = 0.5$ corresponds to the allowed region quoted in ref.~\cite{Brustein:2004jp}.
\begin{figure}[!htb]
\includegraphics[width=8cm]{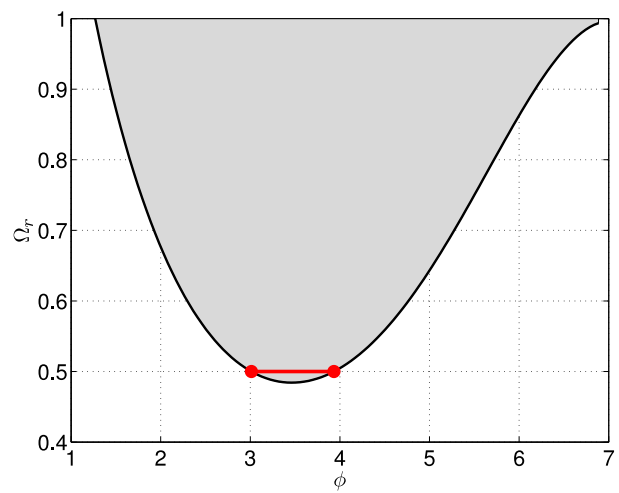}
\caption[BAM]{\label{BAM} Fraction of the total initial energy needed in radiation, as a
function of  the initial value of $\phi$, for the field to end up in its minimum. The solid line along $\Omega_r = 0.5$ denotes the result obtained in
ref.~\cite{Brustein:2004jp} and explained in the text.}
\end{figure}

\subsection{Two field evolution}
\label{sec:2field}
Let us now introduce $\sigma_i$, the imaginary part of the scalar field $\sigma$ in the evolution.
The result of solving the full system given by eqs.~(\ref{evKKLT}) is presented in figure~\ref{re+im},
where we show contour plots for the initial values of the pair ($\phi$, $\sigma_{i}$), that lead
to both fields ending up at their minimum. The different subplots show different contributions of
the background (once again given by radiation).
\newlength\lfig
\lfig=4cm
\begin{figure}[!hb]
\includegraphics[width=\lfig]{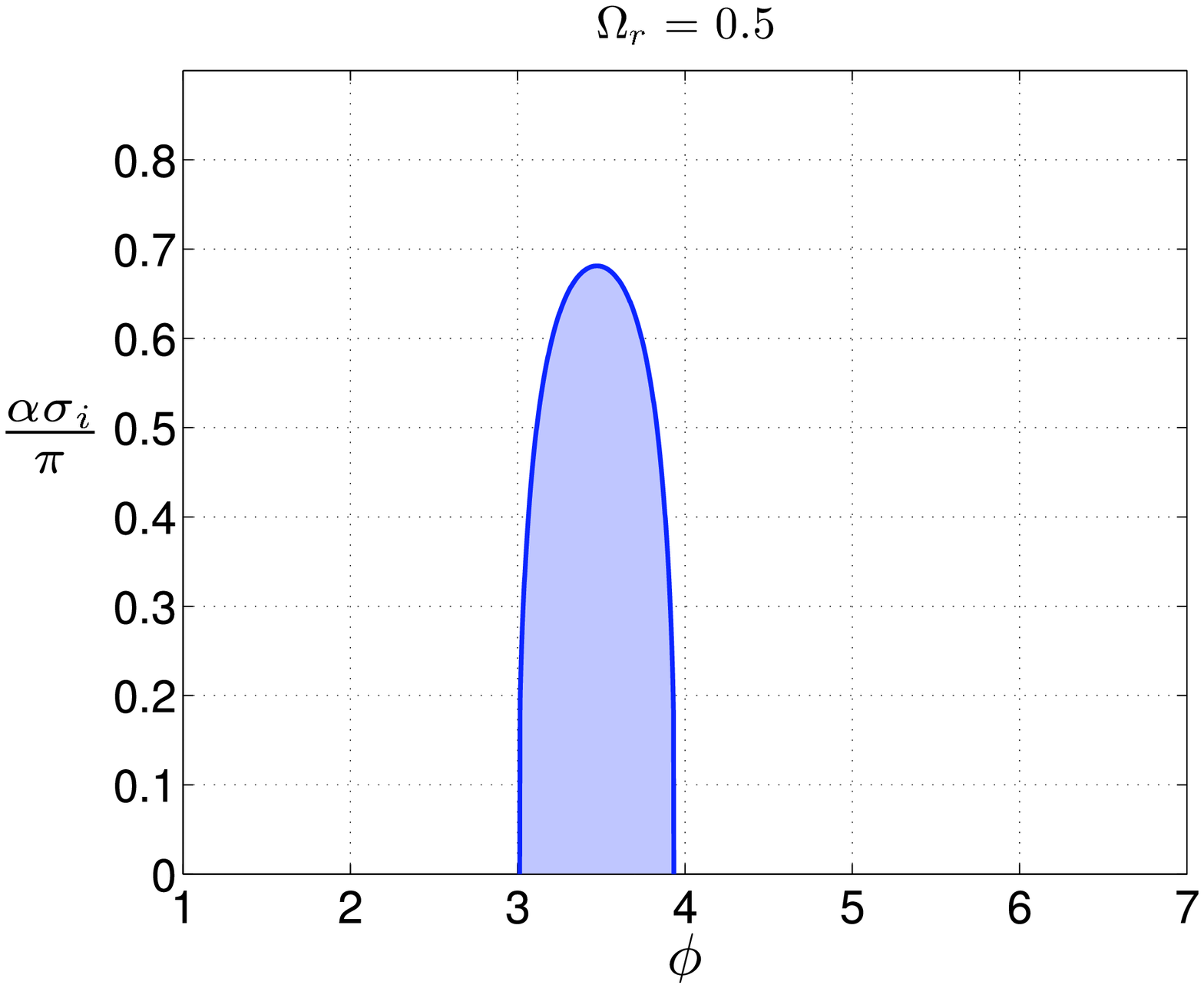}
\includegraphics[width=\lfig]{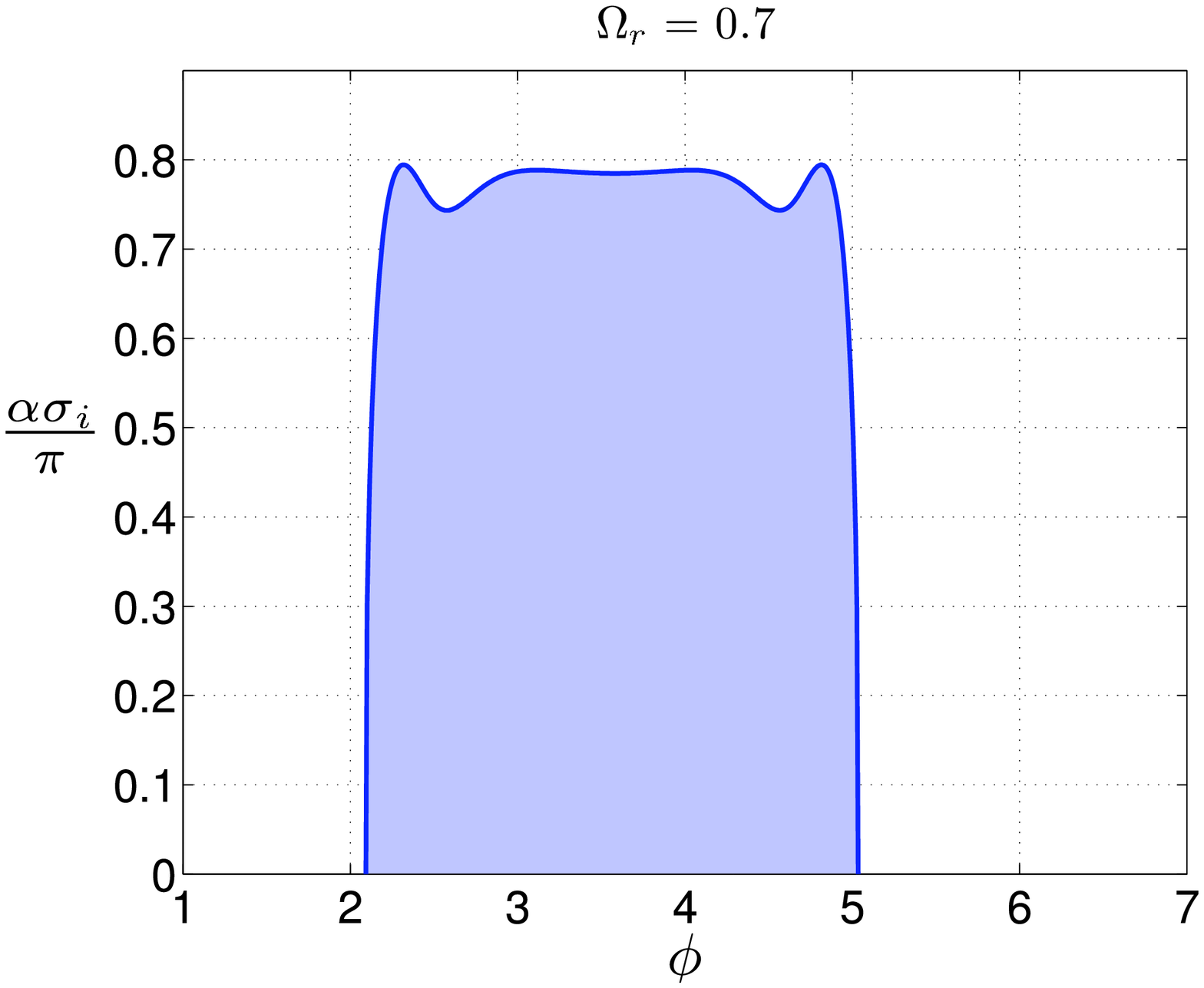}
\includegraphics[width=\lfig]{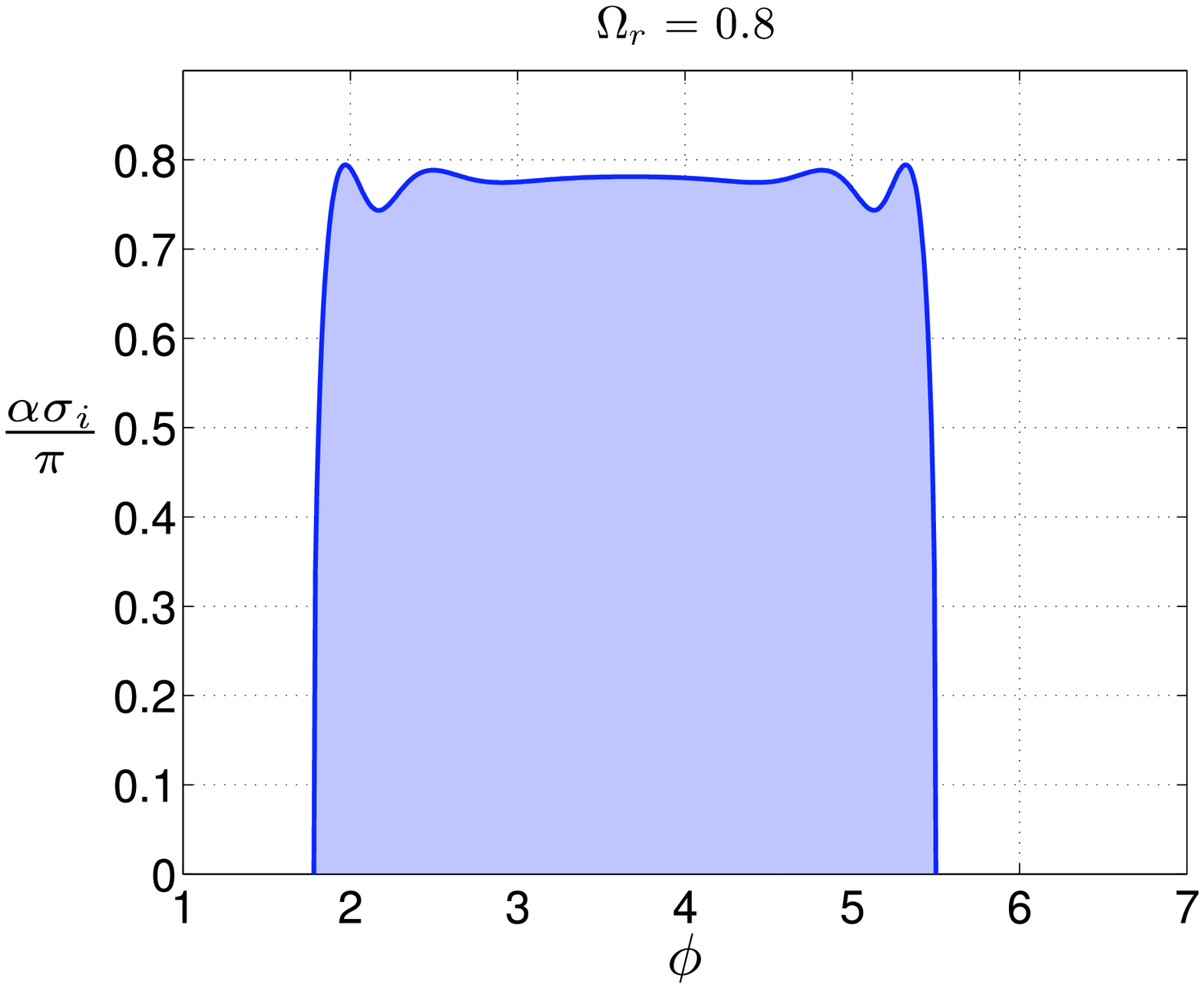}
\includegraphics[width=\lfig]{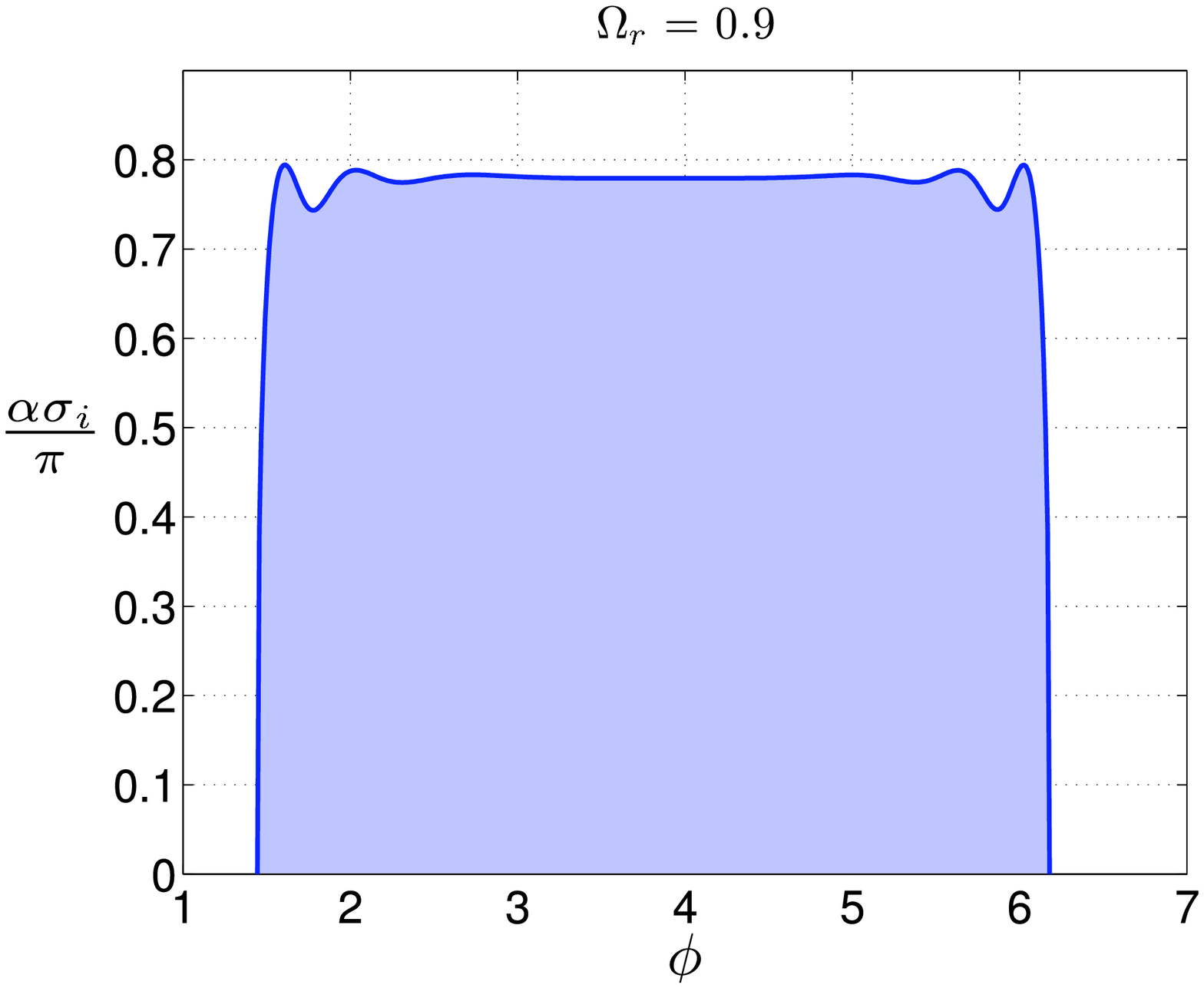}
\caption[re+im]{\label{re+im} Regions where initial values for $(\phi,\sigma_i)$ will lead to a stabilisation of the field in the minimum of the potential, with a radiation background.
The corresponding initial value for $\Omega_r$ is indicated on each subplot.
The plots are symmetric in $\sigma_i$ so we only plot the positive values. The parameter values used in the potential are detailed in the text.
}
\end{figure}
\begin{figure}[!hb]
\includegraphics[width=\lfig]{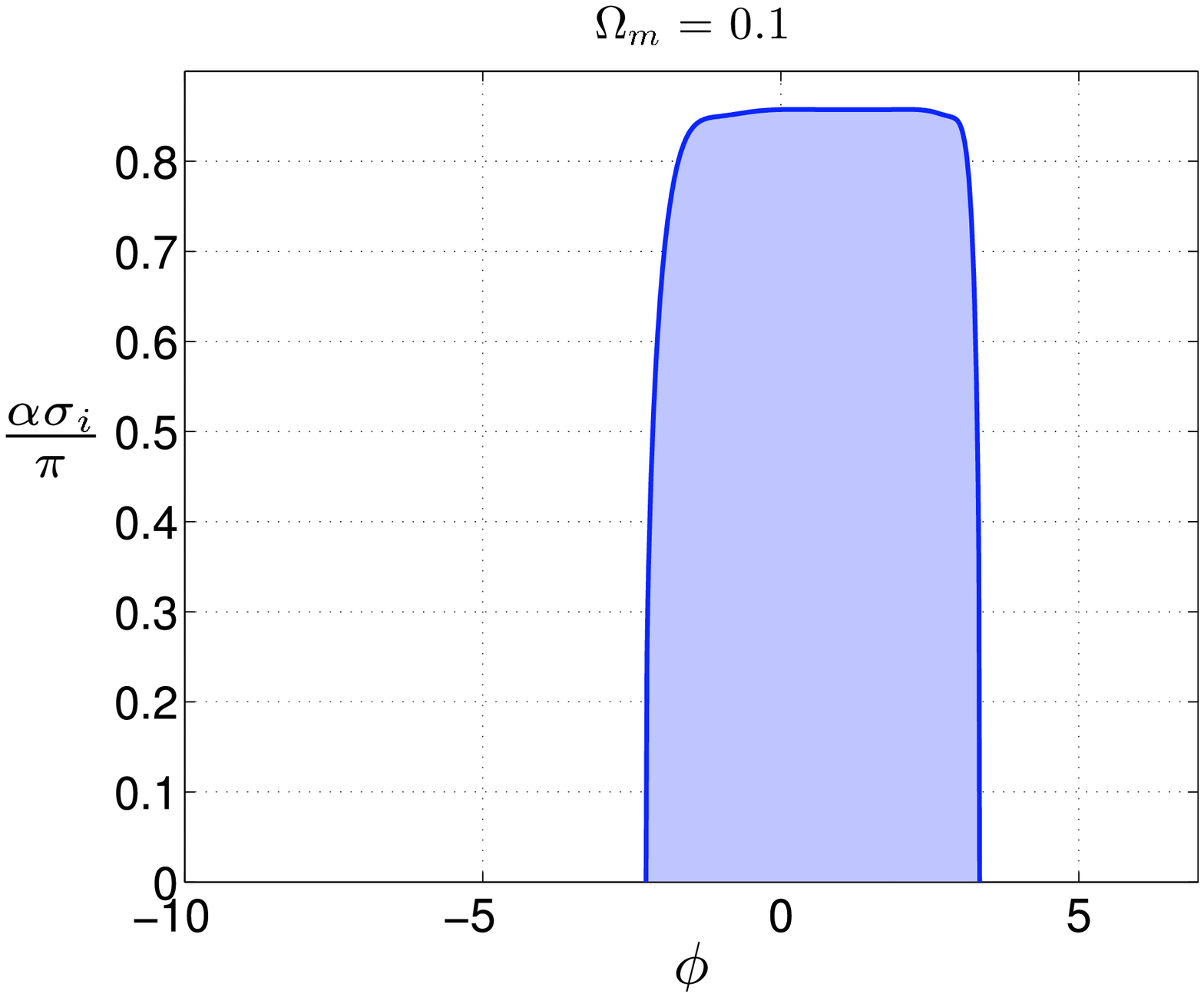}
\includegraphics[width=\lfig]{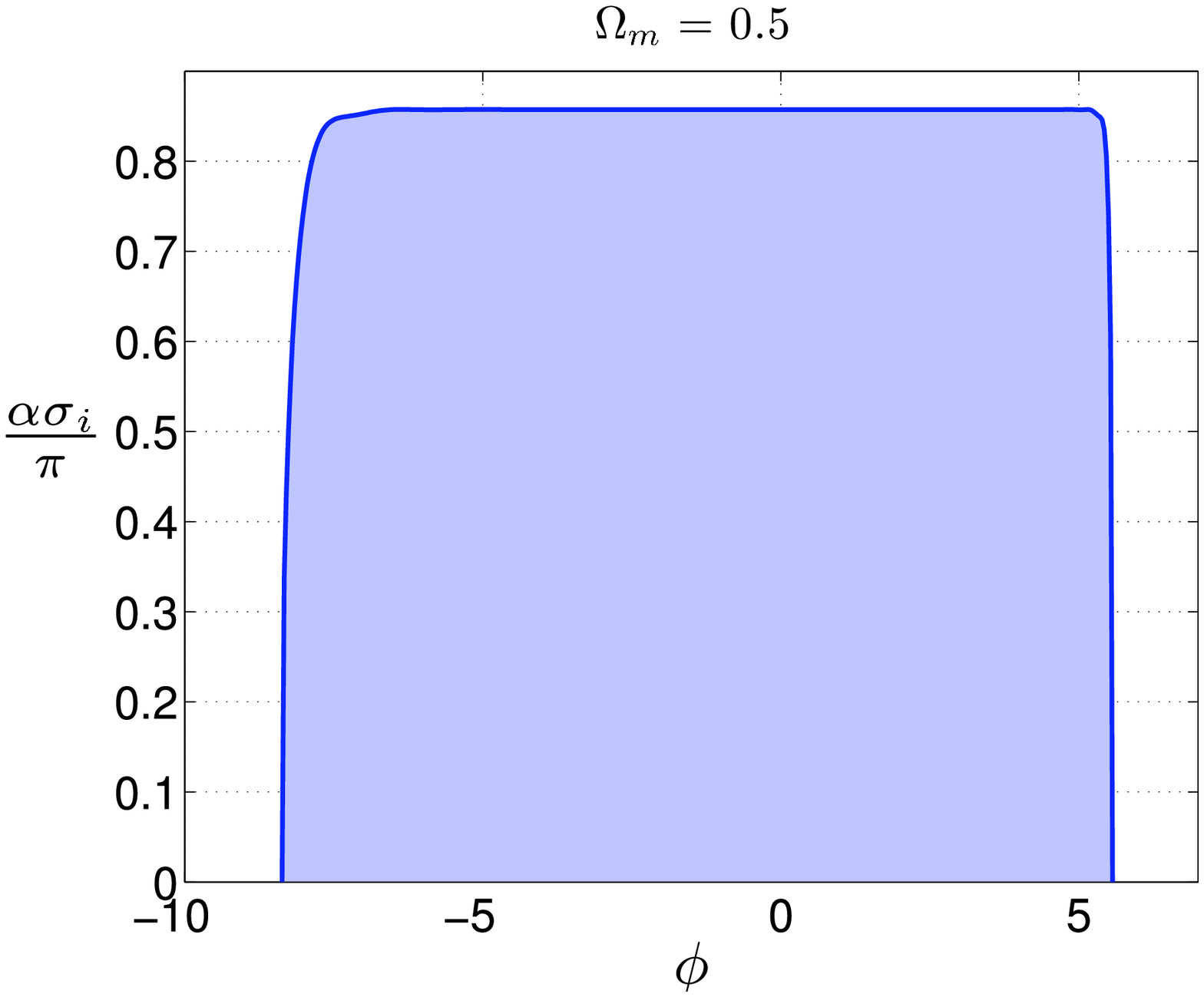}
\includegraphics[width=\lfig]{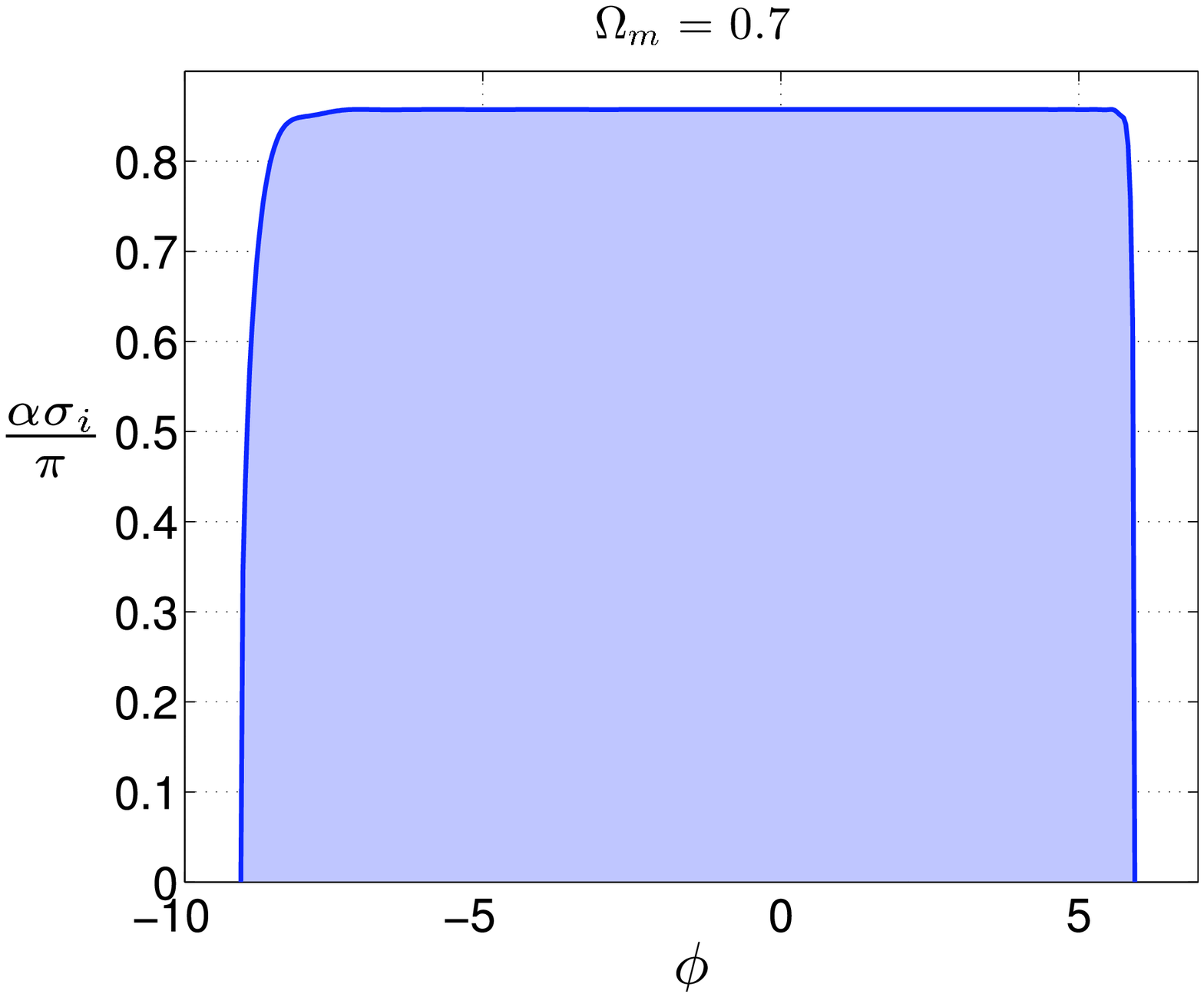}
\includegraphics[width=\lfig]{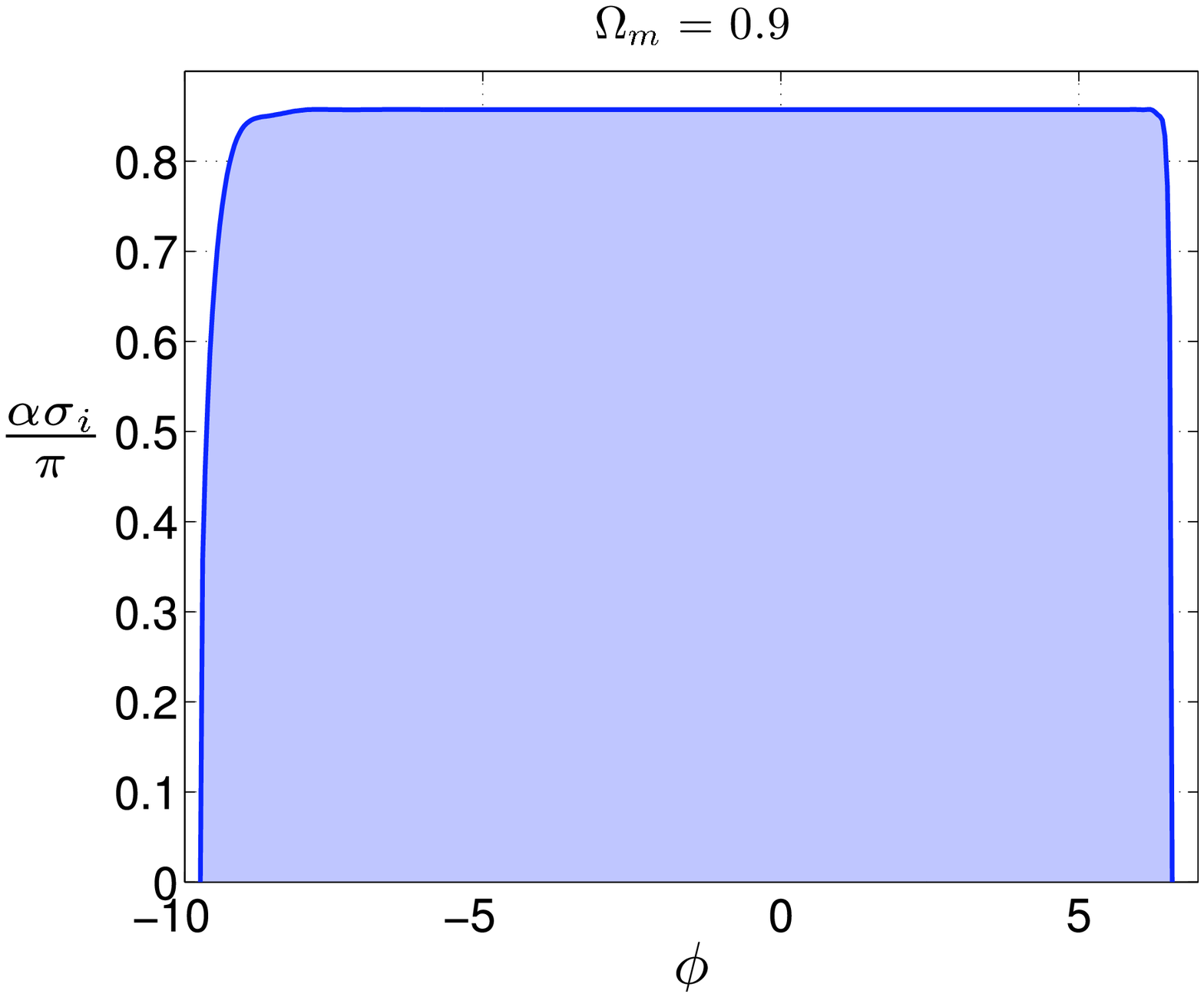}
\caption[fig6]{\label{re+im-mat}
Stabilization regions for the initial conditions $(\phi,\sigma_i)$ with a matter background. Initial value of $\Omega_m$ is shown for each subplot. The potential is the one used in figure~\ref{re+im}.}
\end{figure}

\begin{figure}[!hb]
\includegraphics[width=\lfig]{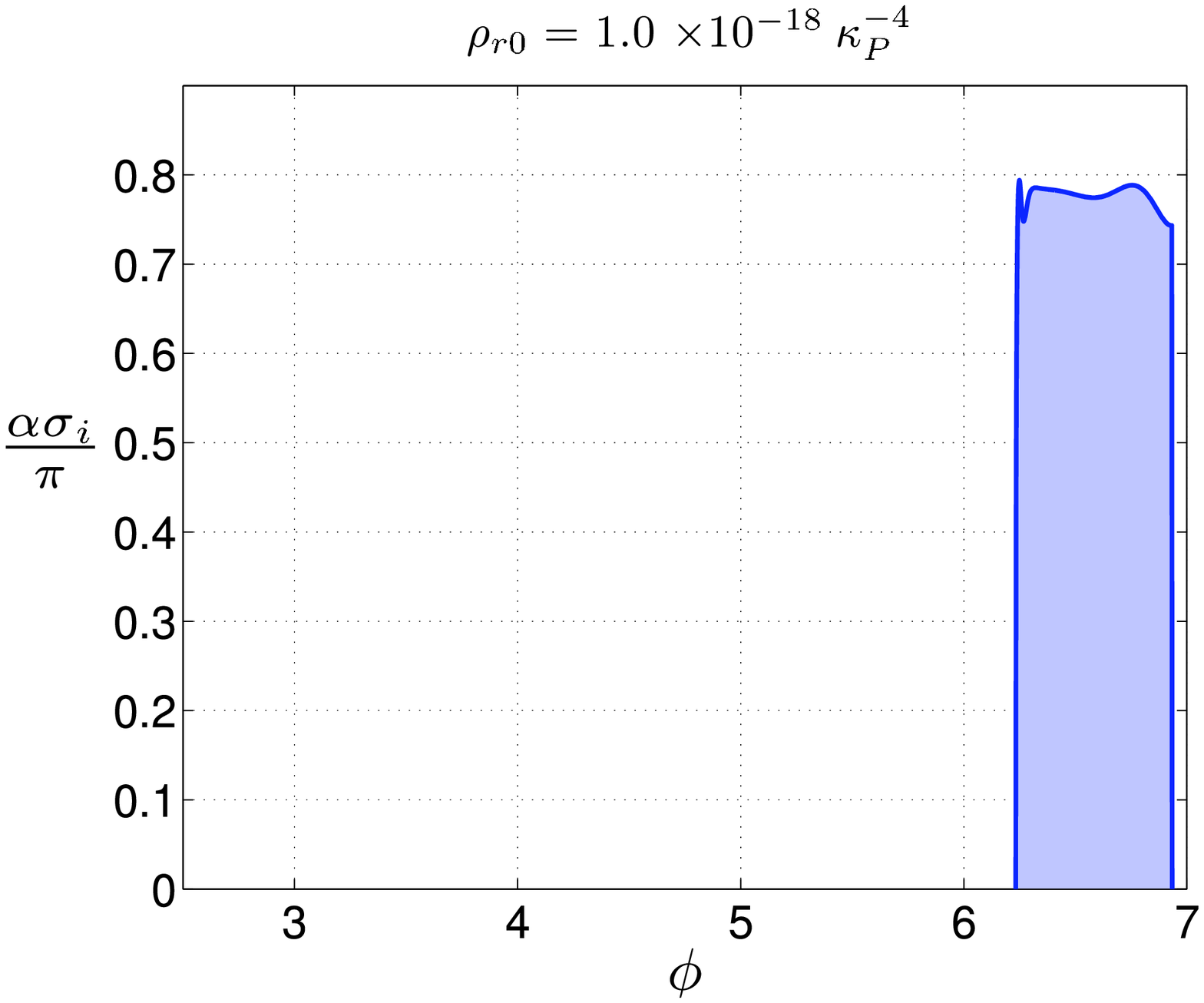}
\includegraphics[width=\lfig]{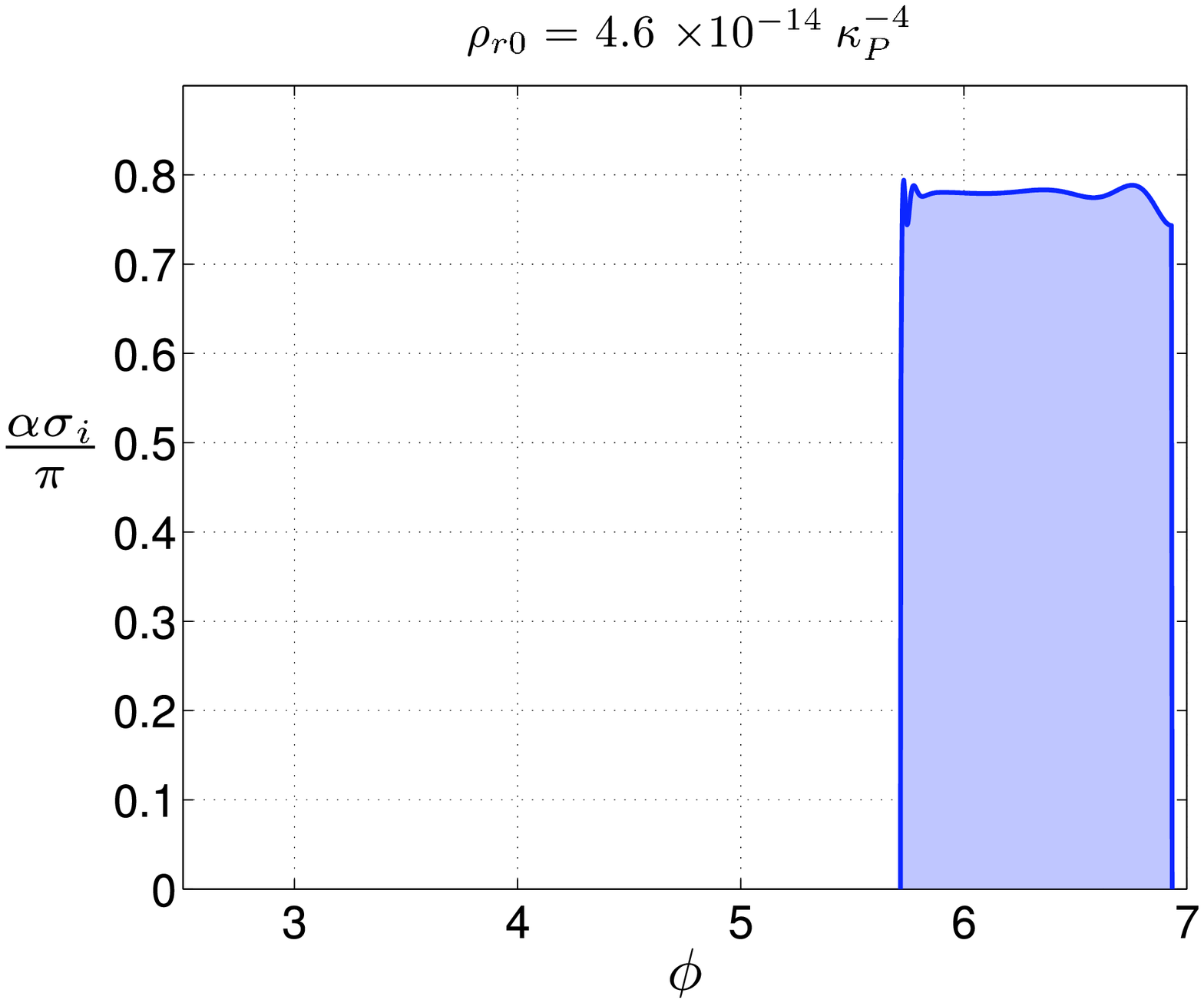}
\includegraphics[width=\lfig]{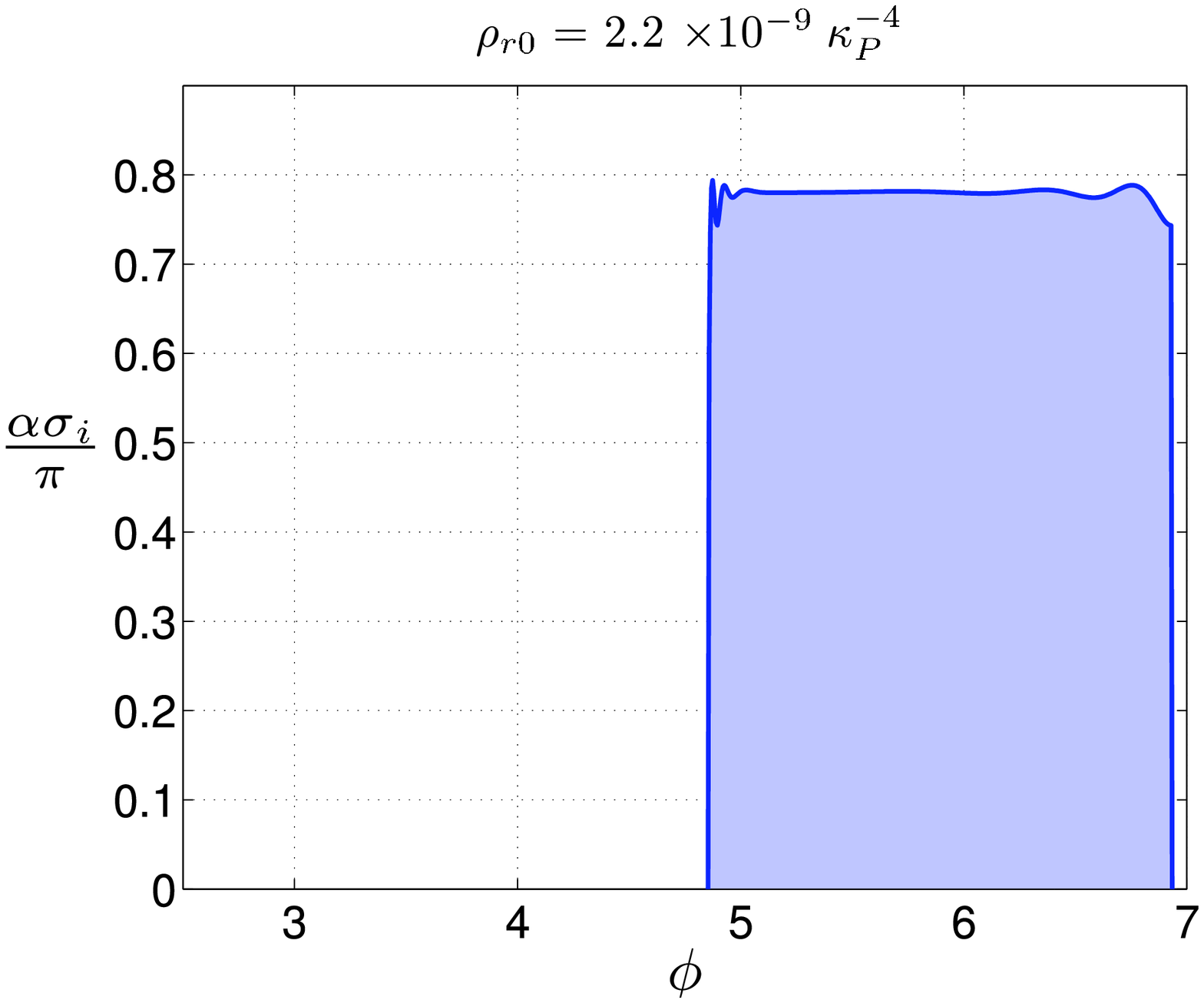}
\includegraphics[width=\lfig]{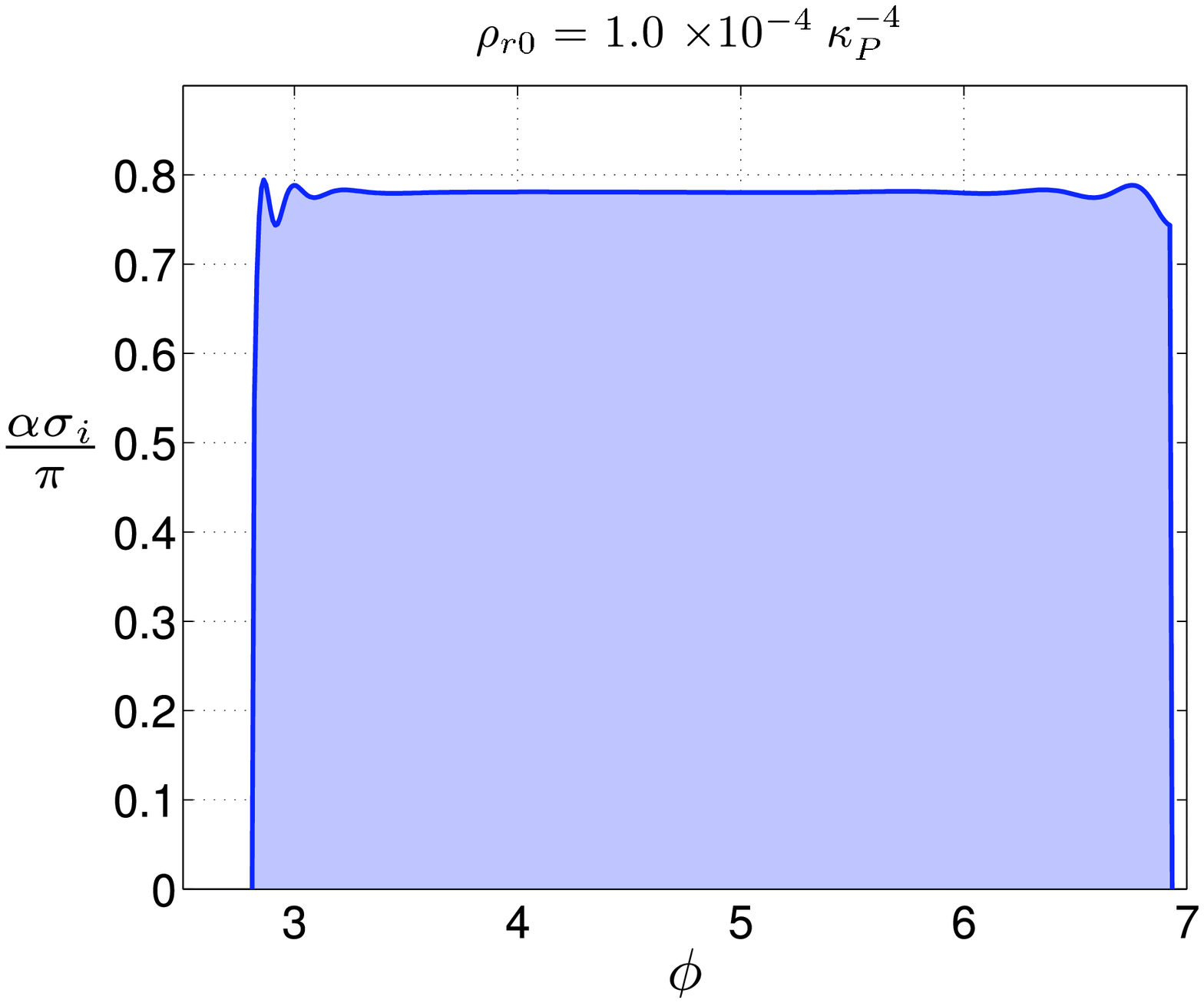}
\caption[fig7]{\label{evrho}
Same results as in figure~\ref{re+im}, but now fixing the initial value of $\rho_r$ instead of fixing $\Omega_r$. The potential is the same.
}
\end{figure}

A few comments are in order.
As can be seen from figures~\ref{BAM}, \ref{re+im} and \ref{re+im-mat},
there is both a minimum and a maximum value of $\phi$ for which stabilisation at the minimum of the potential is successful.

Let us begin with the left bound on the value of $\phi$.
This comes from the
type of potentials we are considering not being exactly an exponential potential. More precisely, they
deviate most from an exponential shape for smaller values of $\sigma_r$.
In these regions a scaling solution does not exist, and
the evolution attractor is a scalar field dominated solution (the reasons for this are explained in more detail in Appendix \ref{apa}).
Consider a typical evolution like the one in figure~\ref{fig20}. Initially the field will have values that correspond to a scalar field dominated
(region 1) evolution. Even if  the background dominates initially (as in figure~\ref{fig20}), the field will eventually take over. In other words, there is a saturation point for the initial value of $\rho_b$, above which all evolutions will be similar (they only differ in a time shift, corresponding to the time during which the field remains frozen initially). After some time the scalar field dominated solution becomes unstable and the field's energy density becomes kinetically dominated (region 2 in figure \ref{fig20}). The background energy density will then dominate and the field will freeze at some constant value (region 3 in figure~\ref{fig20}).
Note that the larger the initial energy density for the field (that is, the smaller the initial value of $\phi$), the longer it will take for the background to freeze the field. For very small initial values, the field will run past the minimum \emph{before} being frozen by the background.
The limiting case corresponds to the field freezing to a value nearly coincident with the position of the minimum (the case discussed in ref.~\cite{Brustein:2004jp}).
A matter ($\gamma = 1$) type of background, having a slower redshift evolution, is much more effective at dominating the total energy density and freezing the field's evolution. This can clearly be seen in a much smaller left bound on the initial values of $\phi$ (see figure~\ref{re+im-mat}).

The bound arising on the maximum value for $\phi$ is also related to the scaling solution.
Naively one would think that, as the initial value approaches the minimum, the field would acquire less kinetic energy and it would be easier to stabilize its evolution in the minimum.
The reason we still get a bound on the maximum allowed initial value of $\phi$,
is really due to
the choice we made for the initial conditions.

If the initial energy density for the field is higher than the scaling one, the field will have to reach the scaling solution
in a kinetic dominated regime, and it will be easy for it to overshoot the minimum if it starts too close to it.
As can be seen from figure~\ref{fig20}, for this type of potentials the
scaling solution is essentially background dominated, so $\Omega_b$ is close to unity.
By choosing to fix the value of $\Omega_b < 1$ (which we did
following ref.~\cite{Brustein:2004jp}) we immediately force the initial conditions for the scalar field to be away from the scaling solution. 
Furthermore, since the potentials are not exact exponentials, 
the scaling value actually increases and goes assymptotically to 1 as we approach the minimum.
Therefore, for most initial values of $\Omega_b$, the field will start above the scaling solution, leading to a right bound on the initial values of $\phi$. This bound will shift towards the minimum as we increase the initial $\Omega_b$.

Of course, eventually, as $\Omega_b$ approaches 1 we can get the right bound to coincide with the position of the minimum.
This is much easier to see if  we choose to fix the initial value of $\rho_b$, instead of $\Omega_b$. In this case, the right bound always coincides with the position of the maximum in the potential (just before the ``roll over'' point). The results are shown in figure~\ref{evrho}, where the right bound is always the same for  changing values of the initial $\rho_b$.

Given the dependence of the scalar potential on ${\rm cos}(\alpha \sigma_i)$ these plots are symmetric
under the inversion $\sigma_i \rightarrow -\sigma_i$ (we only plot the positive $\sigma_i$ region). We can see that up to $85\%$ of the parameter space in the imaginary direction will lead to a stabilization of the field at the minimum.

\section{Kallosh-Linde model}
\label{sec:KLmodel}
This model generalizes the original version of the KKLT model by admiting one additional componet
in the superpotential. More explicitly, we write
\begin{equation}
W = W_0 + A e^{-\alpha \sigma} + B e^{-\beta \sigma} \,,
\label{WAB}
\end{equation}

\begin{figure}[!htb]
\includegraphics[width=8cm]{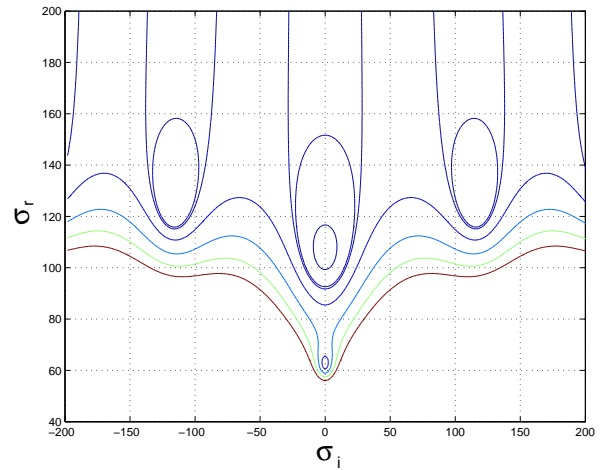}
\caption[contKL]{\label{contKL} Contour plot of the scalar potential, defined by eq.~(\ref{WAB}), in the
($\sigma_r$, $\sigma_i$) plane. The values of the parameters are taken from ref.~\cite{Kallosh:2004yh}, see text. All minima are supersymmetric and AdS, apart from the one at $\sigma_i=0$ and smallest
value of $\sigma_r$, which is Minkowski.}
\end{figure}

The particular example considered in ref.~\cite{Kallosh:2004yh}, sets parameters, $A = 1$, $B = -1.03$, $C = 0$, $\alpha = 2 \pi/100$, $\beta = 2\pi/99$ and $W_0$ such that there is a supersymmetric minimum with zero cosmological constant, i.e. $W_0$ is such that both $W$ and $F_{\sigma} \equiv K_{\sigma} W+W_{\sigma}$ vanish at some $\sigma_r=\sigma_{\rm crit}$, $\sigma_i=0$. Furthermore,
there are a series of supersymmetric, AdS minima. All those are shown in figure~\ref{contKL},
where we plot the scalar potential for this model, in the ($\sigma_r$, $\sigma_i$) plane. The Minkowski
solution is one of the two at $\sigma_i=0$, and corresponds to the smallest value of $\sigma_r$.

\begin{figure}[!htb]
\includegraphics[width=\lfig]{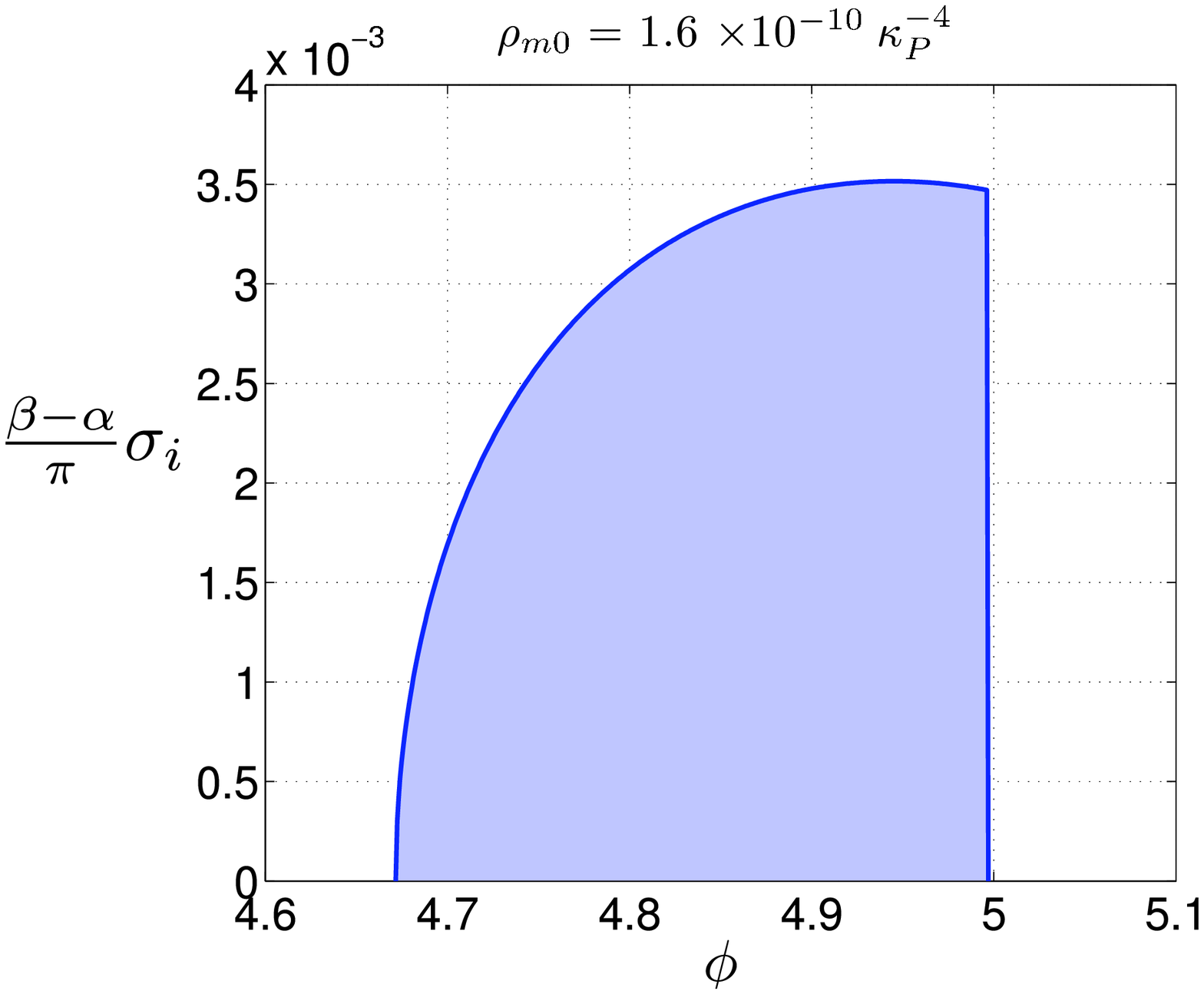}
\includegraphics[width=\lfig]{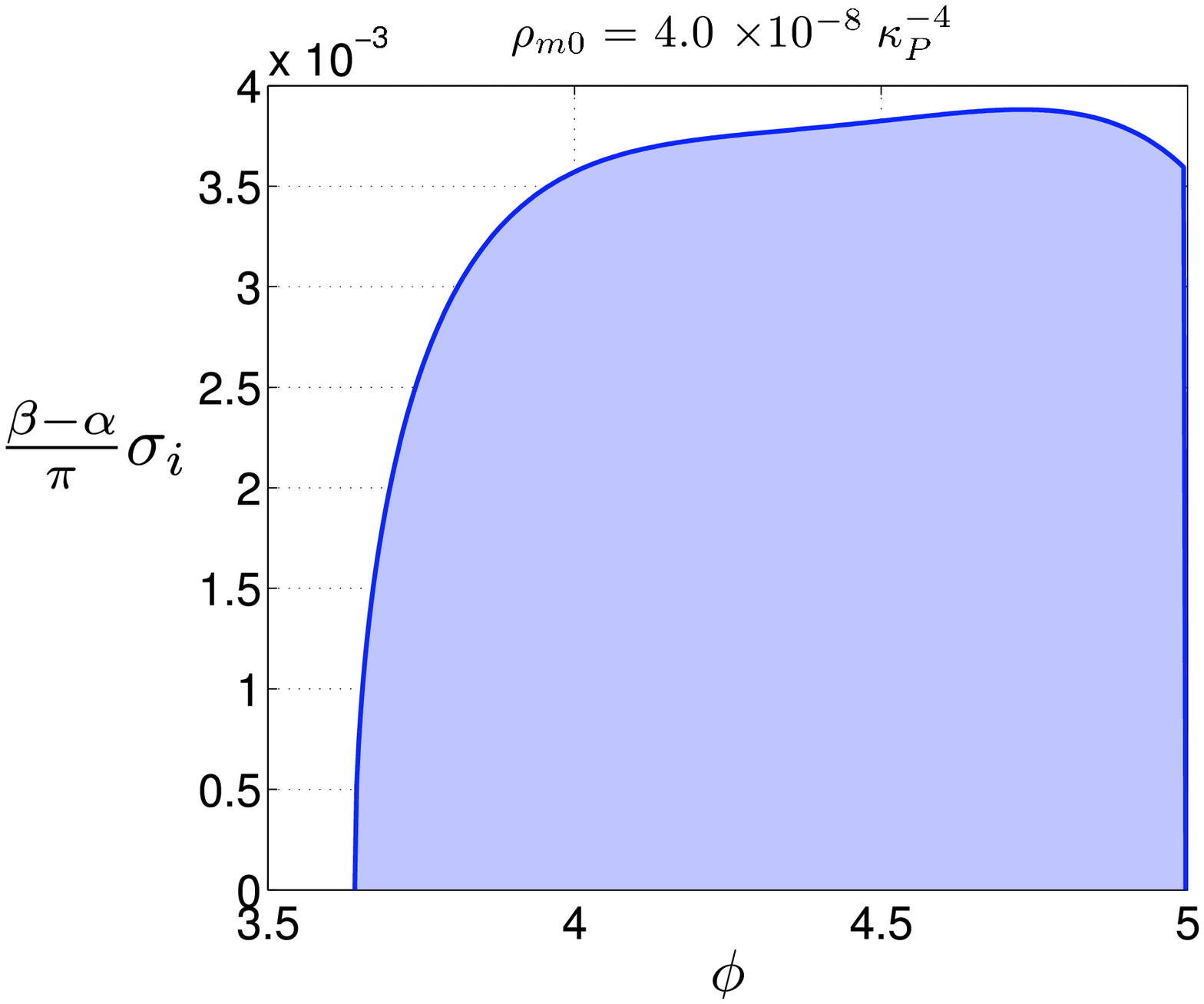}
\includegraphics[width=\lfig]{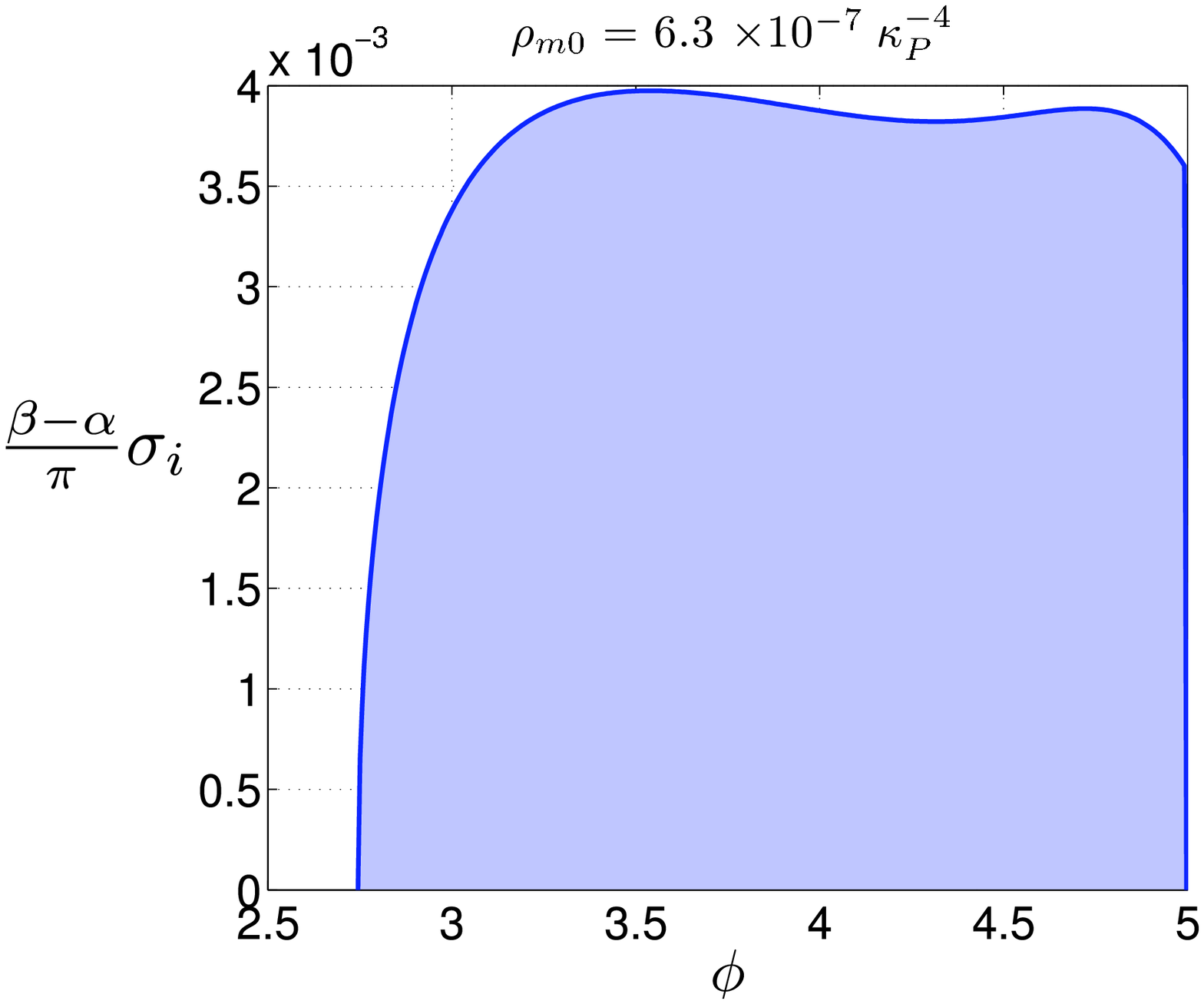}
\includegraphics[width=\lfig]{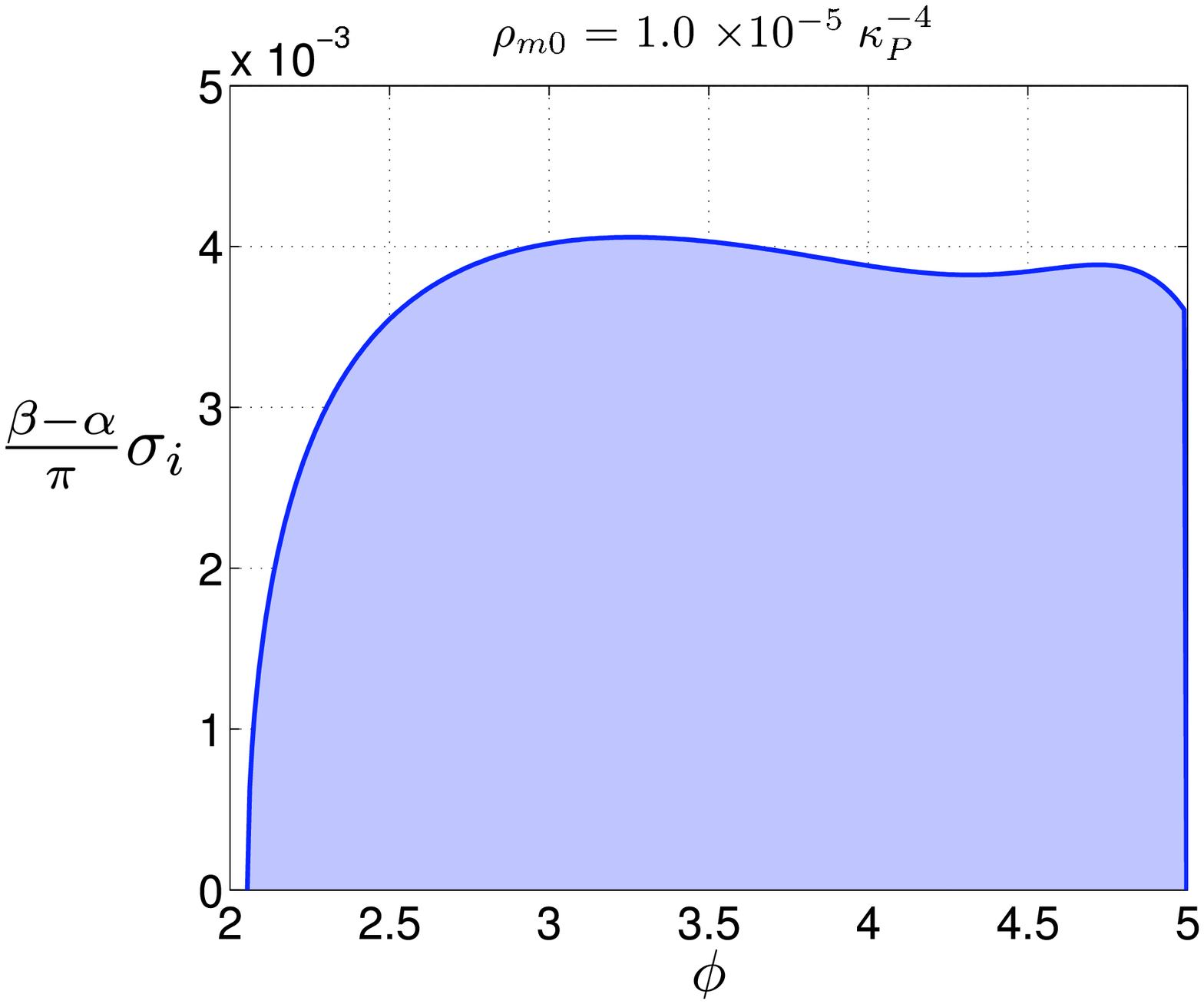}
\caption{\label{L2mat} 
Stabilisation region for the initial conditions $(\phi, \sigma_i)$, with a matter background. The corresponding initial value of $\rho_m$ is shown on each subplot. We used the model of ref.\cite{Kallosh:2004yh}, described in the text after eq.~(\ref{WAB}).
}
\end{figure}

Unfortunately, this choice of parameters makes it almost
impossible for the field to be stabilized if the initial value of the field is far from the minimum. In figure~\ref{L2mat} we show the region of stabilization for a matter background, using a fixed $\rho_m$ for initial conditions.
The reason for this is, again, related to the fact that the potential is very shallow for
$\phi < \phi_{\rm min} \approx 5.08$, hence the scaling solution is never the dominant attractor.

Nonetheless, the mechanism of stabilising the field with a background fluid becomes impressive for models where the minumum lies within the range, in $\phi$, of the scaling solution. One can see this by moving the minimum to larger values of $\phi$, for example, by decreasing $B$ to $ B = -1.5$.

\begin{figure}[!htb]
\includegraphics[width=\lfig]{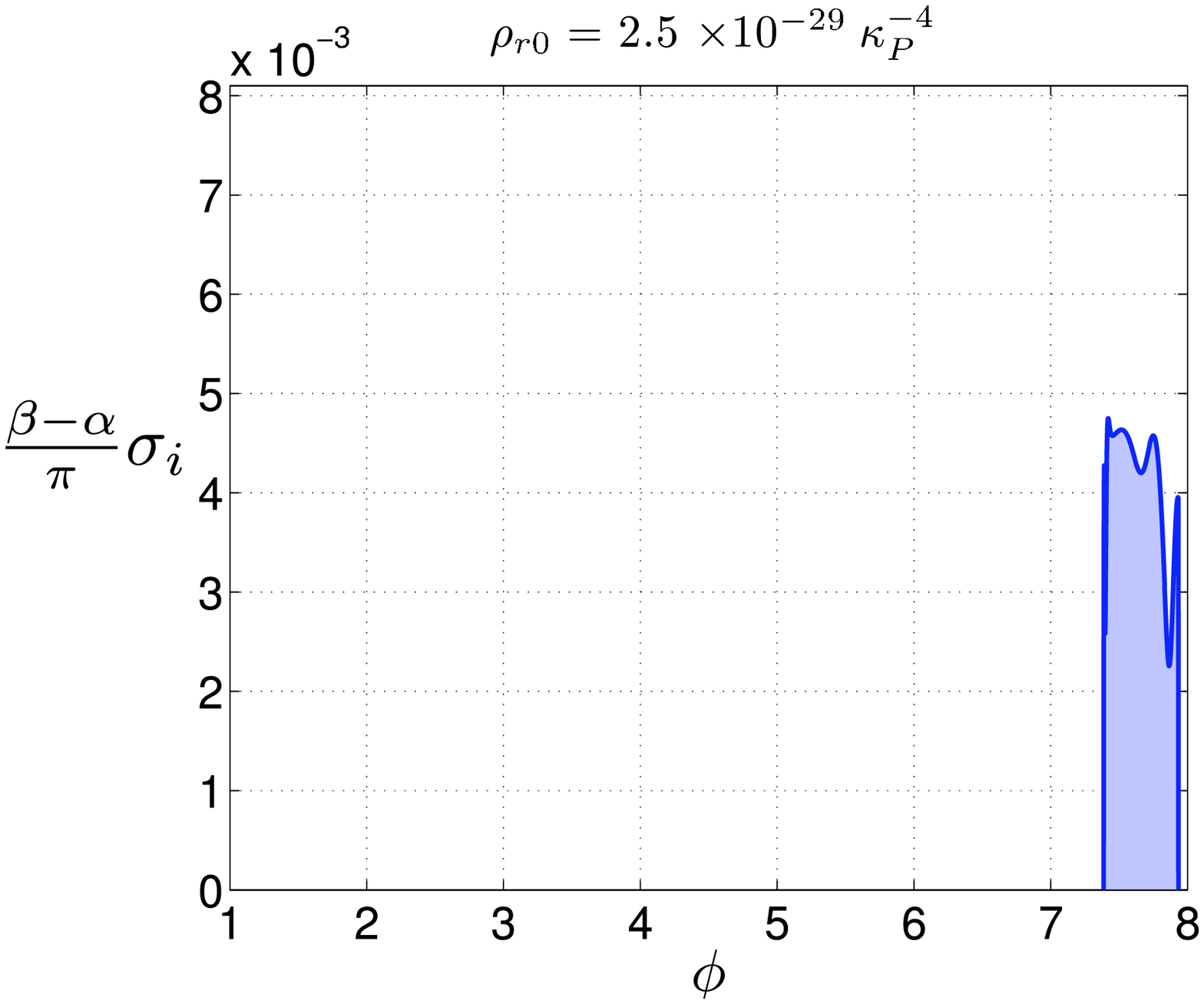}
\includegraphics[width=\lfig]{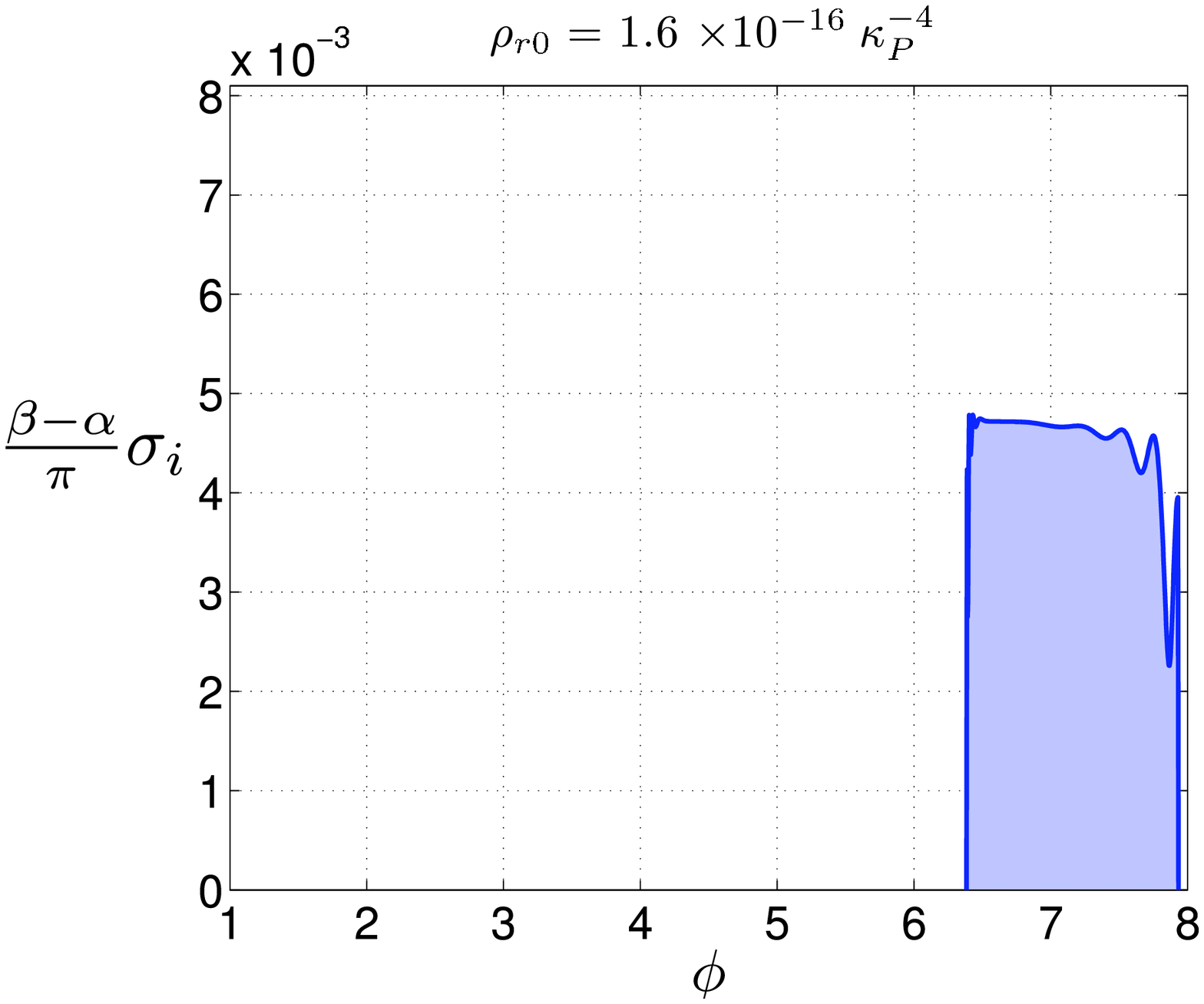}
\includegraphics[width=\lfig]{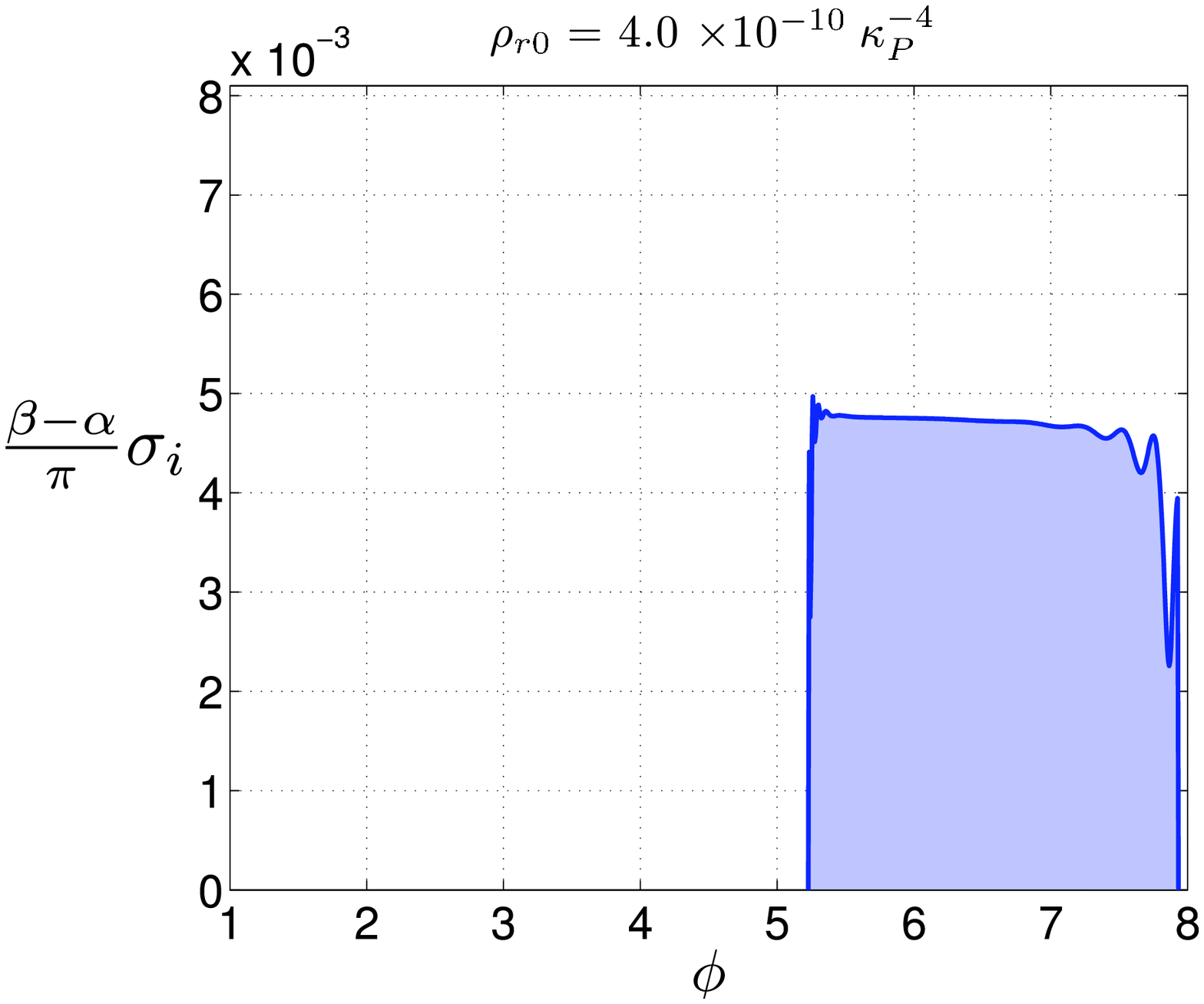}
\includegraphics[width=\lfig]{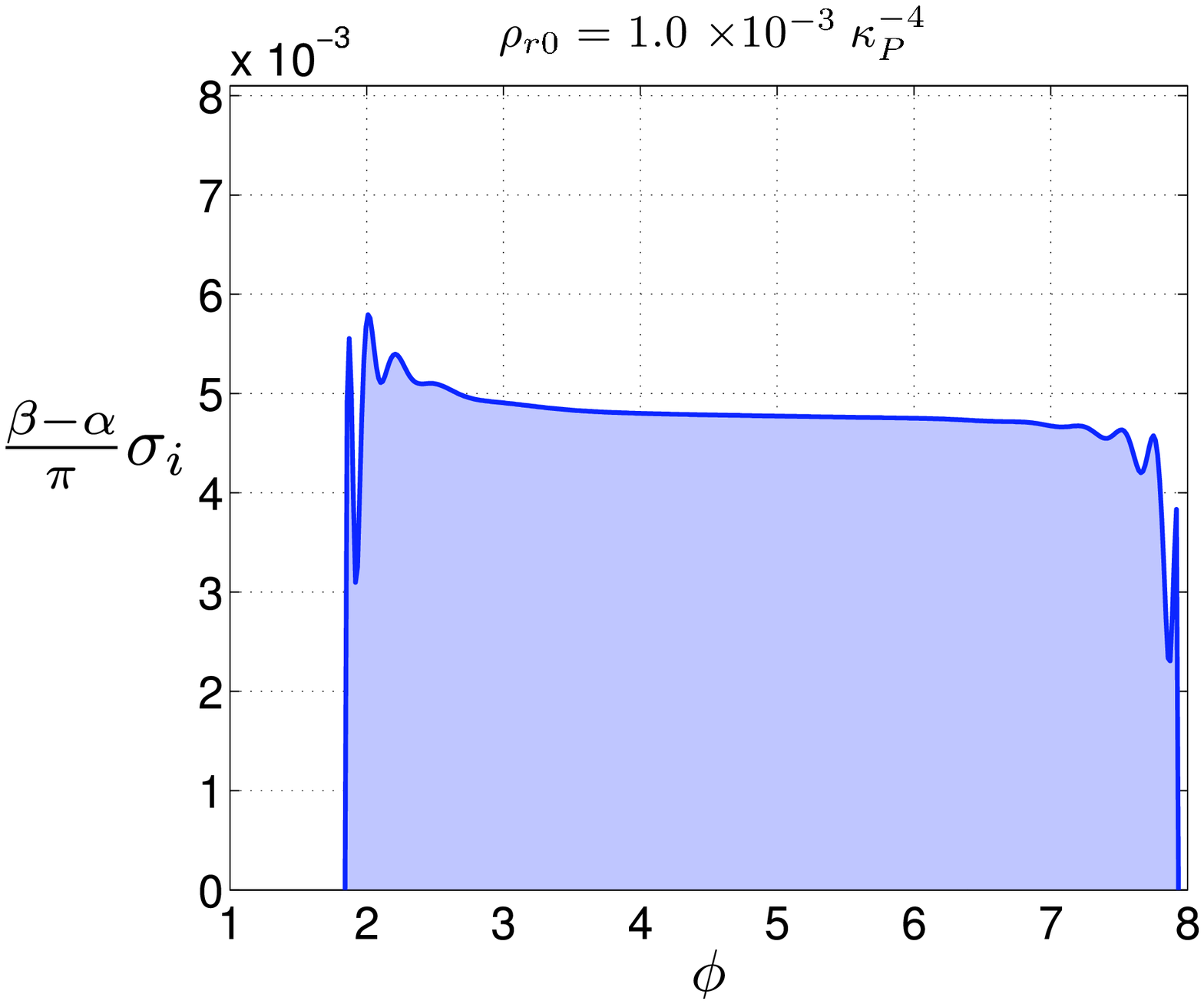}
\caption[fig13]{\label{N1rad}
Region of the initial values in $(\phi,\sigma_i)$ for which stabilisation at the minimum occurs in the model of ref.~\cite{Kallosh:2004yh}, with parameter $B$ set to $-1.5$. We used a radiation background, fixing the initial value of $\rho_r$.
}
\end{figure}

\begin{figure}[!htb]
\includegraphics[width=\lfig]{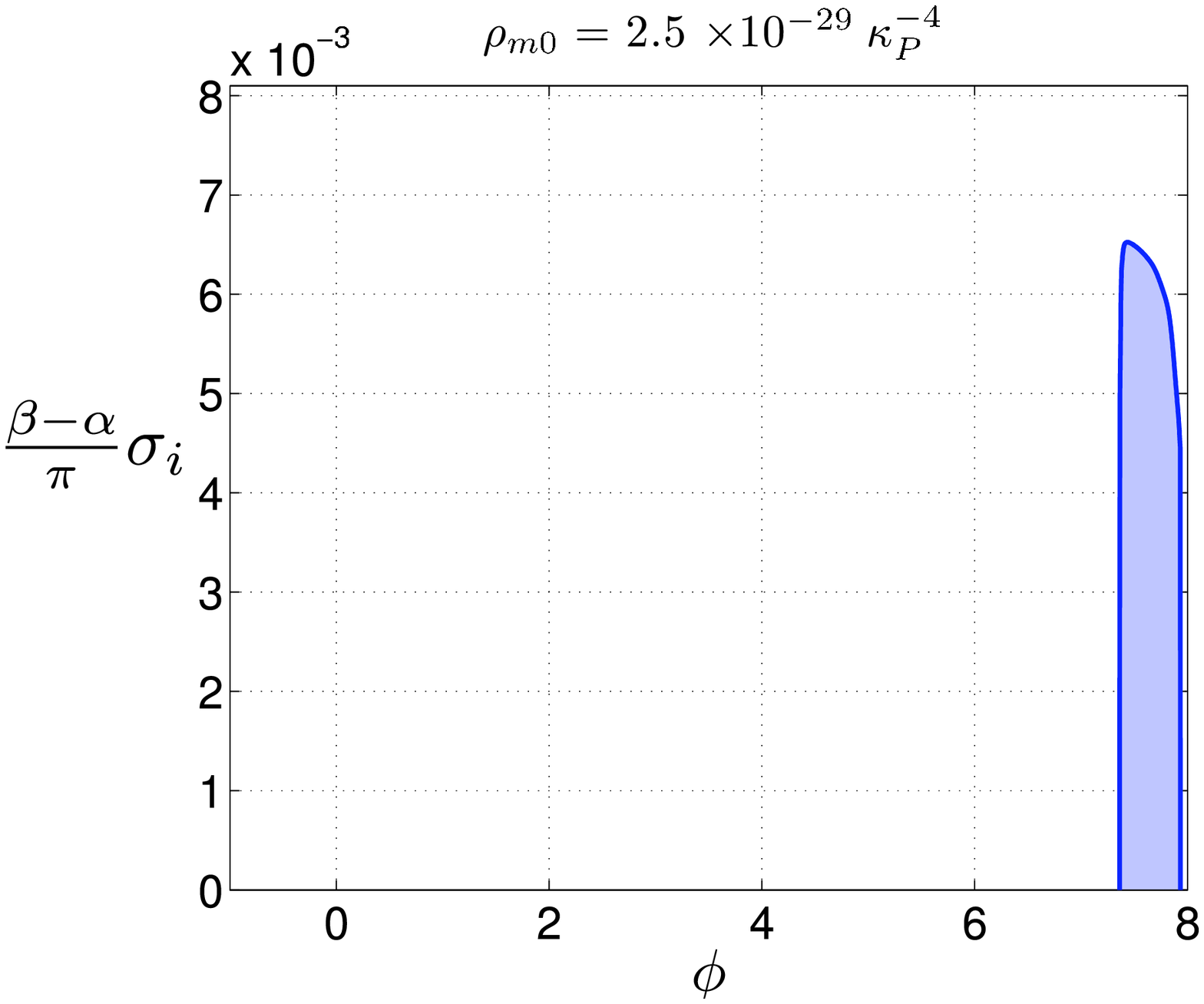}
\includegraphics[width=\lfig]{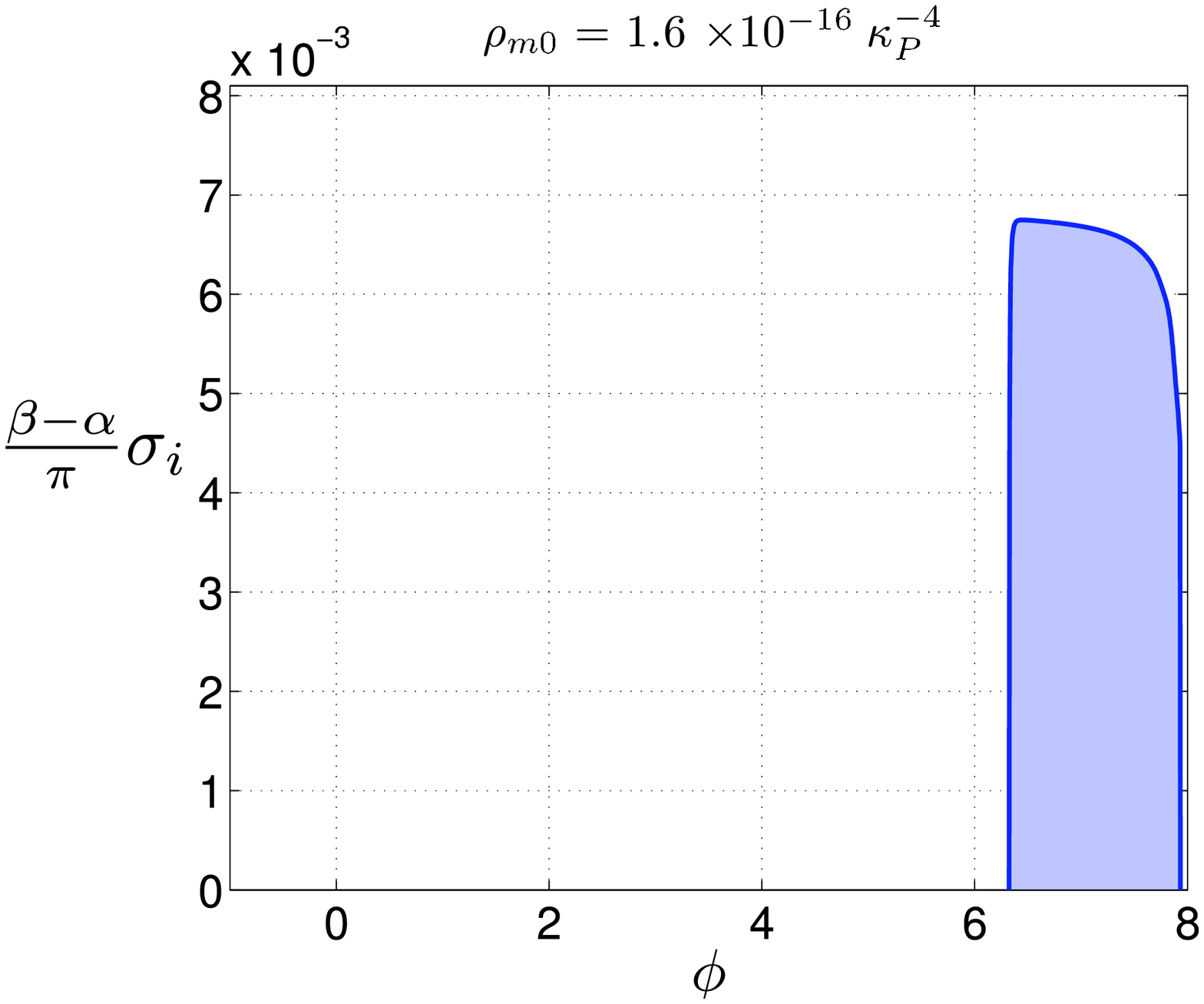}
\includegraphics[width=\lfig]{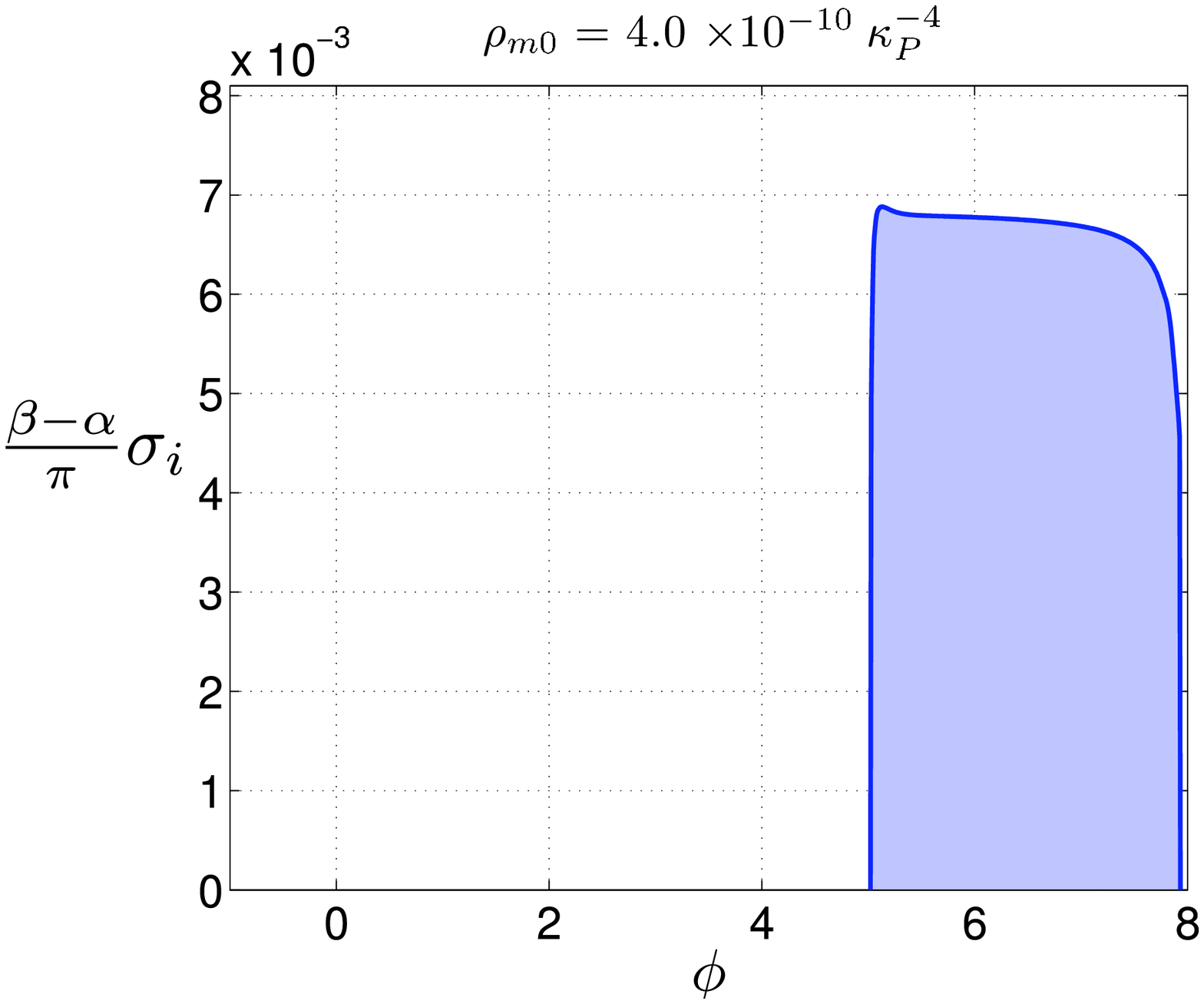}
\includegraphics[width=\lfig]{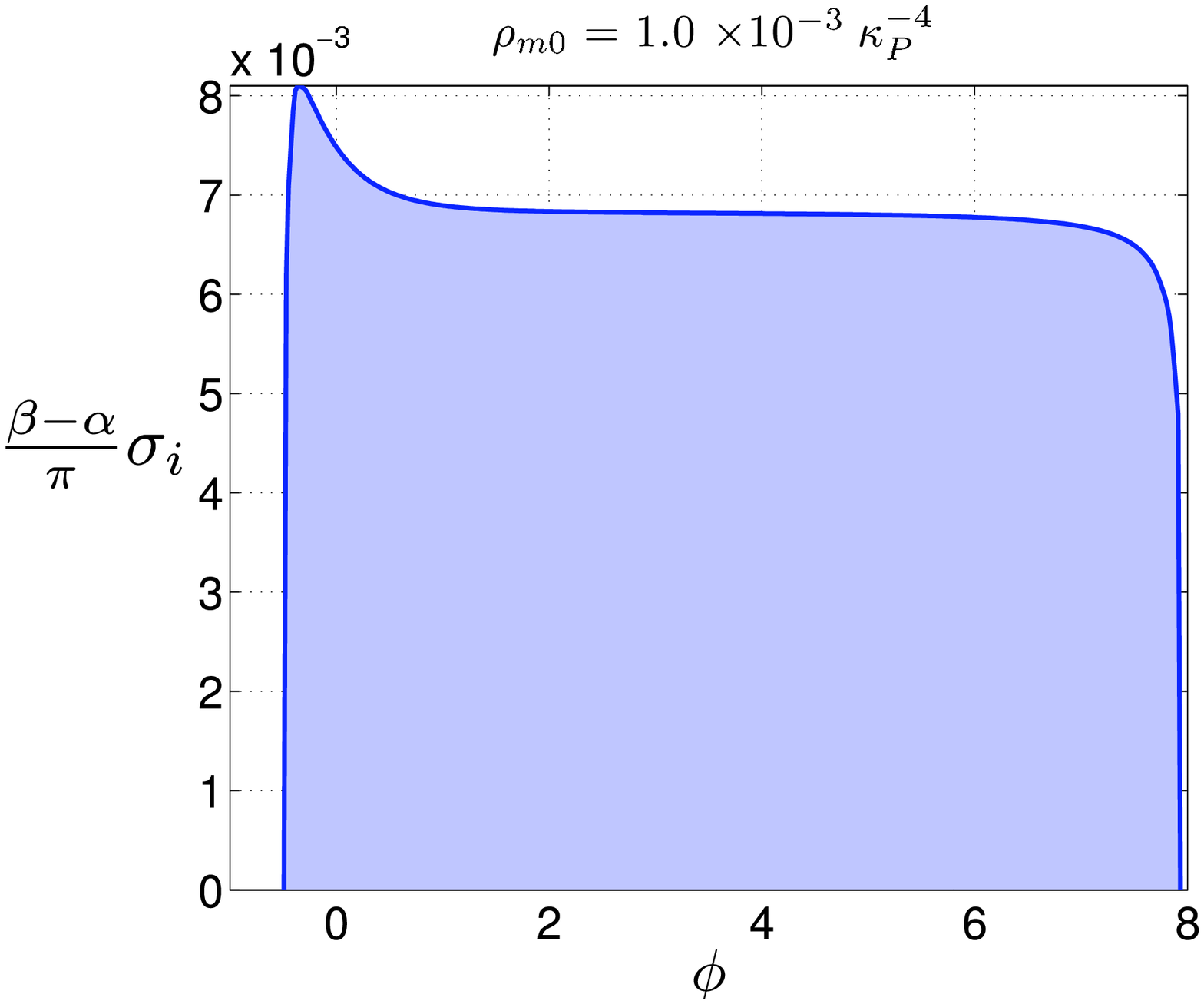}
\caption[N1mat]{\label{N1mat} 
Region of the initial values in $(\phi,\sigma_i)$ for which stabilisation at the minimum occurs in the model of ref.~\cite{Kallosh:2004yh}, with parameter $B$ set to $-1.5$. We used a matter background, fixing the initial value of $\rho_m$.
}
\end{figure}

Following the approach of the previous section,  in figures \ref{N1rad} and \ref{N1mat} we show the bounds on the initial position of the
fields that lead to successful stabilisation. As in the examples of sections \ref{sec:1field} and \ref{sec:2field}, the region of allowed initial conditions increases with the amount of initial energy density in the background. Also, as before, a smaller background equation of state will lead to better stabilization.

In the imaginary direction, this type of potential is much more difficult to stabilize than the ones used in the previous section. For these examples, we can only cover around $0.6 \%$ of the available range of initial conditions. The range of values for $\sigma_i$ are similar in both cases, so this is only due to the fact that the examples in this section have a much larger period in the imaginary direction. This can be easily  checked by comparing figures~\ref{contKL} and \ref{contB}.



\section{CONCLUSIONS}
\label{sec:conclusions}

In this article we have studied moduli evolution in the context of the so-called KKLT-like models,
that arise out of Type IIB string theory, where fluxes and the effect of anti-D3 or D7 branes have been taken into account. We have mainly looked at the allowed region of initial conditions of the moduli fields which leads to their eventual stabilisation at the supersymmetric minima of the scalar potential. 
Employing an approach first developed in~\cite{Barreiro:1998aj}, we found that a background perfect fluid has the ability to slow down the fields preventing them from running past the minimum,  in agreement with~\cite{Brustein:2004jp}. 

Moreover, we have extended the work of ref.~\cite{Brustein:2004jp} to include the dynamical evolution of the imaginary part of the moduli fields. We have evaluated the region of allowed initial conditions for the fields and confirmed that it increases for larger values of the initial background energy density
$\rho_0$ and lower values of its equation of state.
The effectiveness of this mechanism is also dependent on the existence of a scaling solution as the attractor. Regions in which the potential is considerably shallower than an exponential lead to a scalar field dominated attractor, working against successful stabilisation in the minimum.

In the second part of this paper we have turned our attention to analyse the Kallosh-Linde model of
ref.~\cite{Kallosh:2004yh}. The conclusions are qualitatively equivalent but we note that
only a small interval of initial conditions are allowed in the imaginary direction, making it more difficult to obtain stable moduli solutions in this case.

%
%

\appendix
\section{}\label{apa}

In this appendix we summarise the key results
presented in \cite{Barreiro:1998aj} and \cite{Ng:2001hs}, and describe how they can be applied to the discussion in the main text.

In the first of these papers \cite{Barreiro:1998aj}, we considered the issue of stabilising the dilaton in the context of superstring cosmology without the explicit use of non-trivial three form fluxes, but including gaugino condensates and external sources such as radiation.
Making use of the fact that in these models the scalar potentials arising out of gaugino condensates are generally exponential in nature, we demonstrated how scaling solutions associated with exponential potentials \cite{wetterich,ferreira,CLW} could be slightly modified for these models, but basically had similar outcomes. In particular we showed how, due to the friction of the background expansion, the energy density in the scalar field could become, for a period, a fixed fraction of the background density , allowing the field to be trapped in the minimum of its potential, generally as it left its scaling regime. The key point was that previous scaling solutions available for pure exponential potentials were also approximately valid for quasi-exponential potentials. This was further reinforced in  \cite{Ng:2001hs}, where a phase space analysis of  generic quasi-exponential potentials was carried out, looking for scaling solutions.

Consider a scalar field $\phi$, with canonical kinetic terms, evolving in a FRW Universe containing  a fluid with barotropic equation of state $p_\gamma = (\gamma -1) \rho_b$.
 It is useful to analyse the system in terms of the new variables $x\equiv \frac{\dot{\phi}}{\sqrt6 H}$ and $y\equiv \frac{\sqrt V}{\sqrt3 H}$, as they allow for the determination of the scaling solutions. The evolution equations can then be written as \cite{CLW,Ng:2001hs}
\begin{eqnarray}
\label{eq-Hprime}
H' &=& - {\frac{3}{2}} H  [2x^2 + \gamma(1-x^2-y^2)]  \\
\label{eq-x}
 x' &=& -3x +\lambda \sqrt{\frac{3}{2}} y^2 + \frac{3}{2} x  [2x^2 + \gamma(1-x^2-y^2)] \\
\label{eq-y}
y' &=& - \lambda \sqrt{\frac{3}{2}} xy + {\frac{3}{2}} y  [2x^2 + \gamma(1-x^2-y^2)]  \\
\label{eq-lamb}
 \lambda'  &=&  -\sqrt{6} \lambda^2 (\Gamma-1) x\;,
\end{eqnarray} 
where
 a prime denotes a derivative with respect to $N \equiv \ln \frac{a}{a_{\rm initial}}$, and
we defined two new variables \cite{delaMacorra:1999ff,Steinhardt:1999nw},
\begin{eqnarray}
\lambda \equiv - \frac{1}{\kappa_P V} \frac{d V}{d \phi} \,, \hspace{1.5cm}
\Gamma -1 \equiv \frac{d}{d \phi} \left( \frac{1}{\kappa_P \lambda}
\right) \,. 
\end{eqnarray}
For the pure exponential potential, $\lambda$ is a constant and $\Gamma = 1$, so in a sense $\Gamma$ is measuring how much our potential deviates from an exponential one. These equations can be solved easily in a number of regimes \cite{Barreiro:1998aj,Ng:2001hs}.
The types of regimes available were already described in section~\ref{sec:1field} and are shown in figure~\ref{fig20}. Here we will present a more analytic description.

Let us start with the kinetic dominated regime, that is region 2 in figure~\ref{fig20}. In this region, the evolution is dominated by the fields kinetic energy and so the potential energy, $y$, can be neglected in the equations of motion eqs.~(\ref{eq-x}--\ref{eq-y}). This  leads to the solution  \cite{Barreiro:1998aj}
\begin{equation}
\label{eq-xsoln}
x=\left(1+\frac{1-x_0^2}{x_0^2}e^{3(2-\gamma)N}\right)^{-1/2} \;\;,
\end{equation}
where $x_0$ is the initial condition for $x$ (taken at $N=0$, for simplicity). Notice that, since we are neglecting the potential terms ($y$), this solution is really potential independent and is valid as long as we are in a kinetic dominated regime.
Clearly the kinetic energy of the field which is proportional to $x^2$ is decreasing rapidly with time. The solution for $\phi$ follows and is given by 
\begin{eqnarray}
\label{eq-phisoln}
\phi_I(N) &=& \phi_0 + \frac{2\sqrt{6}}{3(2-\gamma)} \left[ \sinh^{-1} \left(\frac{x_0}{\sqrt{1-x_0^2}}\right) \right. \nonumber \\
&-&  \left.
\sinh^{-1} \left(\frac{x_0}{\sqrt{1-x_0^2}}e^{-3(2-\gamma)N/2}\right)\right] \;\;,
\end{eqnarray}
where $\phi_0$ is the initial value of the field. At early times the solution corresponds precisely to that of eq.~(8) of~\cite{Brustein:2004jp}, where the energy density is dominated by the kinetic energy of the scalar field. As time carries on though, and the field slows down, it eventually tends to a constant solution as can be seen in eq.~(\ref{eq-phisoln}). This constant value corresponds to region 3 of figure~\ref{fig20}.

\begin{figure}[!htb]
\includegraphics[width=8cm]{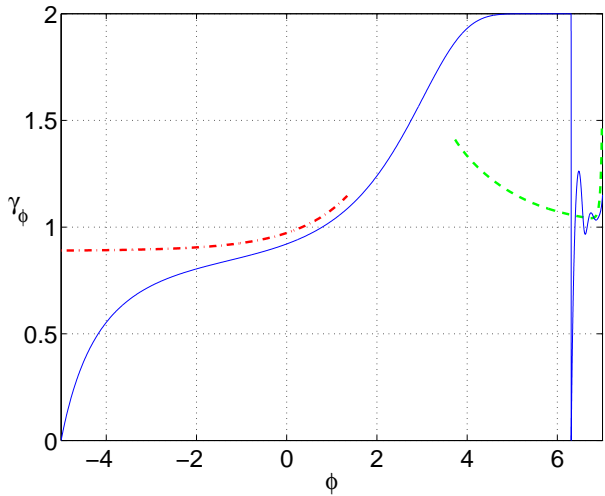}
\caption[fig21]{\label{fig21} Dependence of the scalar field equation of state with the value of the field. Solid line: the numerical integration; dash-dot line: scalar field dominated solution; dash line: scaling solution. The model is the same we used in section \ref{sec:KKLT model} with a background of dust ($\gamma = 1$), i.e. the same as in figure~2.}
\end{figure}

Let us now turn our attention to the scaling evolution, that is region 4 in figure~\ref{fig20}. In  \cite{Ng:2001hs} an approximate equation of state for the field evolution in the scaling regime was derived. This is given by
\begin{eqnarray}
\label{eqstate}
\gamma_{\phi} &=&  \frac{1}{2} \left[\gamma + (2\Gamma -1)\lambda^2/3
\right] \nonumber \\
&-&
 \frac{1}{2} \sqrt{ \left[ - \gamma +
                      (2\Gamma-1)\lambda^2/3 \right]^2
                      + 8\gamma(\Gamma-1)\lambda^2/3 } \,. \nonumber \\
\end{eqnarray}
When $\Gamma \to 1$ (the exponential case), $\gamma_{\phi}$ tends  to $\gamma$, that is one recovers the usual result that the field evolution mimics the background. Moreover, this scaling solution is only stable if it satisfies the following conditions,
\begin{eqnarray}
\label{gfalso}
\gamma_{\phi} &<& \sqrt{\frac{\gamma \lambda^2}{3}} \nonumber
                  \,, {~~~ \rm and}   \\
\gamma_{\phi} &<& \frac{3\gamma-2\lambda^2}{6} \left[
                  1- \sqrt{ 1 + \frac{12 \lambda^2 (2+\gamma)}
                  {(3\gamma-2\lambda^2)^2}} ~\right ] \,.
\end{eqnarray}
In the particular case of double exponential potentials, $V(\phi) = V_0 \exp(a e^{b \phi})$, it is possible to obtain an approximate solution for the field evolution. The most useful way to present it is in a recursive form
\cite{Barreiro:1998aj,Ng:2001hs}
\begin{eqnarray}
\label{sol2}
S_0 &=& \frac{1}{a} \, \ln \left[ \frac{3}{2}
           \frac{\rho_0}{V_0} \frac{\gamma (2-\gamma)}
           {a^2 b^2} \right] \nonumber\,, \\
S_1 &=&  S_0 - \frac{3 \gamma}{a} N  \,, \\
S_k &=& S_0 - \frac{2}{a} \ln (S_{k-1}) -   \frac{3 \gamma}{a} N \nonumber\,,
\end{eqnarray}
where $S \equiv \exp(b \phi)$. It is easy to check that in the regions where the potentials discussed in sections~\ref{sec:KKLT model} and \ref{sec:KLmodel} are close to an exponential in $\sigma_r$, these solutions are a very good fit to the scaling evolution with $S \approx \sigma_r$. 
Finally, there is a third possible type of evolution, shown in region 1 of figure~\ref{fig20}, the scalar field dominated solution. Its effective equation of state was shown in \cite{Ng:2001hs} to be
\begin{equation}
\label{scalareqst}
\gamma_{\phi} = \lambda_{\phi}^2/3 \,,
\end{equation}
where we have defined
\begin{equation}
\label{lambdaphi}
\lambda_{\phi} = \frac{3}{2} \left[\frac{1
                 - \sqrt{1-4(\Gamma-1)\lambda^2/3}}
                                     {(\Gamma-1)\lambda}\right]
\end{equation}
for $\Gamma \neq 1$, and $\lambda_{\phi} = \lambda$ otherwise.
Again, a condition for the stability of the solution was derived, and is given by
\begin{eqnarray}
\label{eigenv2}
6~(\lambda_{\phi}-\lambda)\frac{1}{\lambda_{\phi}}
          + \frac{1}{2}(\lambda_{\phi}^2 + \lambda \lambda_{\phi}
          - 6\gamma)  &<& 0 \,, \\
3 - 6~\frac{\lambda}{\lambda_{\phi}} + \frac{1}{2}
              \lambda_{\phi}(3 \lambda_{\phi} -2 \lambda)\, &<& 0.
\end{eqnarray}
We can now apply this knowledge to the potentials we are dealing with in this paper. In
figure~\ref{fig21} we compare the evolution of the equation of state resulting from a numerical integration with the scaling and scalar field dominated solutions given above, in their region of stability. The model used is the KKLT example presented in section \ref{sec:KKLT model}, also used to produce figure \ref{fig20}. We can clearly see from figure~\ref{fig21} that in this specific example the attractor solution only makes sense for $\phi > 3.7$. Therefore, if the field starts its evolution before this value, it is either in the scalar field dominated solution or in a kinetic energy dominated solution. This makes the stabilisation of the field more difficult to achieve for smaller initial values of the field.


\begin{acknowledgments}
The authors wish to thank Ramy Brustein for  useful
discussions, and Nemanja Kaloper for pointing out the relevance of ref.~\cite{Kaloper:1991mq} to
our work. EJC would like to acknowledge the Aspen Center for Physics for their support when part of this project was started. 
TB is supported by FCT grant SFRH/BPD/3512/2000. BdC is supported by PPARC. NJN is supported by the Department of Energy under contract DE-FG02-94ER40823 at the University of Minnesota. 
\end{acknowledgments}



\end{document}